\documentclass[reprint,aps,pre,amsmath,amssymb,showkeys,superscriptaddress]{revtex4-2}
\usepackage{flushend}
\usepackage{longtable}
\usepackage{graphicx}
\usepackage{dcolumn}
\usepackage{bm}
\begin{document}


\title{Temporal Graphs and Temporal Network Characteristics \\ for Bio-Inspired  Networks During Optimization}

\author{N. DiBrita}
\affiliation{Colgate University, Hamilton, NY 13346, USA}
\author{K. Eledlebi}
\affiliation{Khalifa University, Abu Dhabi 127788, UAE}
\author{H. Hildmann}
\affiliation{Netherlands Organization for Applied Scientific Research, 2597 AK, The Hague, NL}
\author{L. Culley}
\affiliation{Colgate University, Hamilton, NY 13346, USA}
\author{A. F. Isakovic}
\affiliation{Colgate University, Hamilton, NY 13346, USA}
\affiliation{corresponding author email: aisakovic@colgate.edu iregx137@gmail.com}
%
%
%
%

\date{\today}

\begin{abstract}
Temporal network analysis and time evolution of network characteristics are powerful tools in describing the changing topology of dynamic networks. This paper uses such approaches  to better visualize and provide analytical measures for the changes in performance that we observed in Voronoi-type spatial coverage, particularly for the example of time evolving networks with a changing number of wireless sensors being deployed. Specifically, our analysis focuses on the role different combinations of impenetrable obstacles and environmental noise play in connectivity and overall network structure. It is shown how the use of (i) temporal network graphs, and (ii) network centrality and regularity measures illustrate the differences between various options developed for the balancing act of energy and time efficiency in network coverage. Lastly, we compare the outcome of these measures with the less abstract classification variables, such as percent area covered, and cumulative distance travelled.
%
%
\end{abstract}


\maketitle

\section{Introduction, motivation, and research questions}\label{sec:Intro}
Networks that evolve in time, such as infectious disease contact networks, wireless sensor networks, and many others, have received considerable interest in the past two decades \cite{article31,doi:10.1126/science.aai7488, article36}. 
Focus has been placed on
understanding their topology, interruptions (desirable and undesirable) within their structure, their optimization and adaptive operation. Understanding networks often relies on complex multi-parametric and/or multivariate settings. We relied on harnessing the formalized mechanics of emergent behavior based on an our own adaptation (cf. \cite{drones4030033} for details) of the Voronoi tessellation \cite{9300245} which some consider a bio-inspired optimization technique \cite{camazine2003self}.

In addition to a Voronoi-only approach, we used our own hybrid approach, with Genetic Algorithm (GA) hybridized with Voronoi tessellation as the resulting emergent self-organization behavior has been shown \cite{9300245} to be promising for solving coverage problems in realistic model situations. Our technique, termed Bio-Inspired Self-Organizing Networks (BISON), leverages the converged movement towards Voronoi cells’ centers with an intelligent node provisioning algorithm to deliver a fully automated WSN that rapidly self-deploys itself within any finite indoor environment without prior knowledge of the size and structure of the target space \cite{drones4030033, 9300245}.

The BISON algorithm was merged with a localized Genetic Algorithm (GA) to push the trade-off between the pace of space exploration and energy expense further towards faster deployments, especially when faced with complex obstruction structures. 
The mechanics of the approaches are omitted for reasons of brevity but are summarized in SOM (Supporting Online Materials) - A and are available (Open Access) in  \cite{drones4030033}. After considering the inherently dynamic nature of BISON and its GA hybrid variant (GA + BISON), we relied on temporal network analysis as a natural enhancement to the discovery and analysis of changes within the networks generated by these algorithms \cite{https://doi.org/10.1111/1365-2656.12764}.
Voronoi-type algorithms have been utilized for similar purposes, with groups such as Wang et. al 
developing different methods for node reallocation based on Voronoi edges \cite{1624337}, and Zou et. al introducing a spreading algorithm (NSVA) to deploy a fixed number of nodes within an obstacle free region
\cite{10.4108/icst.bict.2014.257917}.

Analyzing the behavior of animals in their environmental networks reveals that they do not randomly interact with each other but rely on temporal properties of their networks \cite{DEB2020106506, https://doi.org/10.1111/1365-2656.12659}. Furthermore, interactions among different animal species are inherently dynamic and change with time and the context of the medium, such as reflections of the outside circumstances \cite{2013tnuc.book...65C, https://doi.org/10.1111/ecog.04716, article06}. 

Thus, performing temporal network analysis on dynamic networks provides information about individual members, the relationship between network nodes, and allows to scale from a few-individuals behavior to a larger scale population level \cite{https://doi.org/10.1111/1365-2656.12764}. Moreover, linking the dynamic patterns of connections in time with the changing status of the members provides insight on the role of the network structure for the specified application.

Centrality and correlations in temporal networks have previously been studied, notably in
\cite{PhysRevE.68.026128, article-g}, where it is shown that correlations indicate the level of connectivity in temporal networks. Naturally, there is a number of reasons motivating the studies of temporal networks, ranging from the need to build surrogate networks \cite{PhysRevE.103.052304}, examining the potential for the emergence of small world networks \cite{PhysRevE.97.062312, PhysRevLett.88.128701, article-h}, to studying networks as model system representing phase transitions, such as \cite{PhysRevE.64.041902} and \cite{PhysRevE.103.062302}.

\pagebreak
In this paper, we demonstrate some of the advantages and trade-offs of BISON (Voronoi-only) and GA-BISON (GA + Voronoi) in a sampling of different environmental conditions, using a temporal network analysis framework employing centrality measures and graph nodes statistics. 

Specifically, we show that network theoretic characteristics and temporal graphs offer a more thorough insight into time evolution of networks than cumulative, application driven measures, such as cumulative distance traveled (CDT), or percent area coverage (PAC)
\cite{drones4030033, 9300245} as well as the effect of simulated noise on the network topology. 

We are targeting the following research questions:
\begin{enumerate}
\item How is the deployment process affecting the connection in the region of interest?
\item What is the temporal difference between the proposed algorithms in terms of the connections achieved between the nodes, over the course of the overall simulation time?
\item How can we quantify the influence of the simulated noise in the on the distribution and the connectivity of the nodes in the network?
\item What do we learn from the time traces of network characteristics, such as regularity and centrality?
\item How do temporal characteristics compare with application-driven measures, such as ADT, PAC?
\end{enumerate}

The paper is organized as follows: section \ref{sec:BG} briefly discusses our Voronoi-only and GA + Voronoi approaches,
section \ref{sec:SotA} provides an overview of recent work related to temporal and static network analysis applied to WSN;
section \ref{sec:BasTN} deliberates the temporal network framework applied to Voronoi-only and GA + BISON;
section \ref{sec:measureRegularity} and
\ref{sec:measureCentrality} focus on the regularity and centrality measures respectively, and
section \ref{sec:Conclusion} concludes the results.


\section{Overview of the BISON and GA-BISON approaches}\label{sec:BG}
We previously reported the progress in autonomous self-deployment of a WSN into two-dimensional bounded target spaces of unknown geometry and topology based on a variant of a Voronoi-based algorithm \cite{9300245}. The proposed method assumed entering the target space from the selected inlet (e.g., doors), and triggering the sequential controlled and optimized release of vehicle or drone carried WSN nodes, which autonomously spread and connect throughout the space to rapidly form a blanket coverage network ready for delivering variety of sensing, monitoring or communication services. 

\begin{figure}[b!]
    \centering
    \includegraphics[width = \hsize, height = 8.2cm]{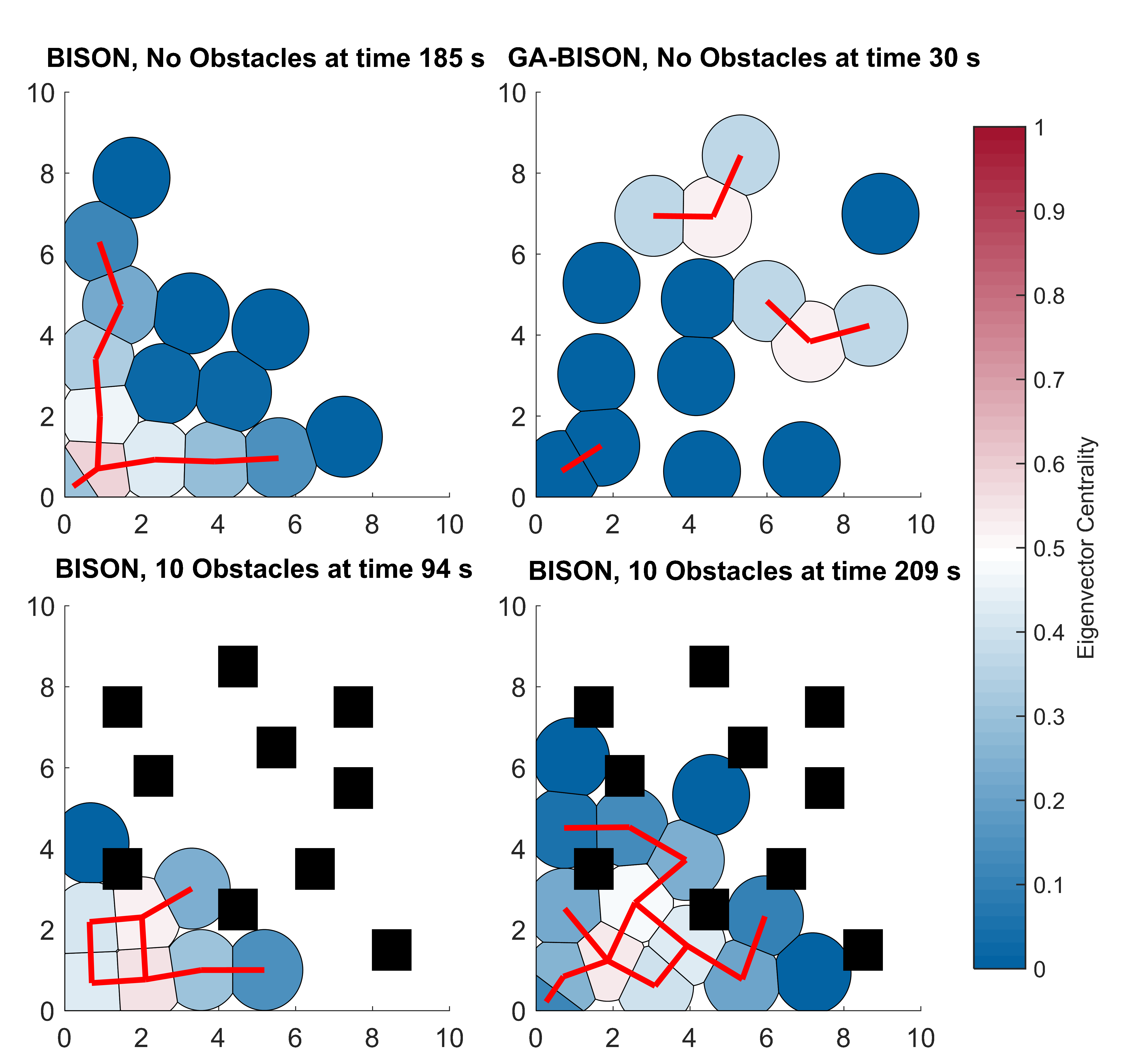}
    \caption{Illustration of the evolution of nodes using Voronoi-only and Ga + Voronoi algorithms. The red lines indicate a suitable network connection for information routing. Both approaches are also capable of expanding into obstacle filled regions. Nodes are colored based on their eigenvector centrality (see section VI for a more detailed discussion). Additional examples of network evolution 
    are offered in SOM - B.}
    \label{fig:Evolution} 
\end{figure}

The sensor nodes autonomously move toward their range-dependent, partially observable Voronoi cells’ centers, as shown in Figure \ref{fig:Evolution}, maintaining a stable collision-free flow designed to rapidly explore and cover the whole target space at the minimum possible time, using as few nodes as possible and draining as little energy as possible, all without any prior knowledge about the geometry of the space and the obstacles in it.

Extensive sets of simulated deployment experiments demonstrate convergence to the stable near-full coverage network, achieved at a fraction of a deployment cost and time, compared to competitive models reported in the literature, which allow to consider BISON as a strong new approach to AI-flavored, blanket-coverage WSN deployment achieved virtually without any human intervention.

In an attempt to further improve, simplify and generalize Voronoi-only deployment, the core BISON algorithm is merged with the localized Genetic Algorithm (GA) applied to push the trade-off between the pace of space exploration and the energy expense further towards faster deployments, especially when faced with complex obstruction structures. The GA with Voronoi approach has an influence on the discovery process of the next best possible local position of nodes in the network \cite{kramer2017genetic}. The opportunity of evaluating several candidate solutions instead of Voronoi centroid allowed sensor nodes to discover better locations, hence enhancing the network performance at faster rates \cite{5872882, 7100294, article23}. GA allowed us to control the randomization of the candidate solutions to best fit our optimization problem, by tuning the location and the rate of the generated solutions in every iteration.

\begin{figure}[h]
    \centering
    \includegraphics[width = \hsize]{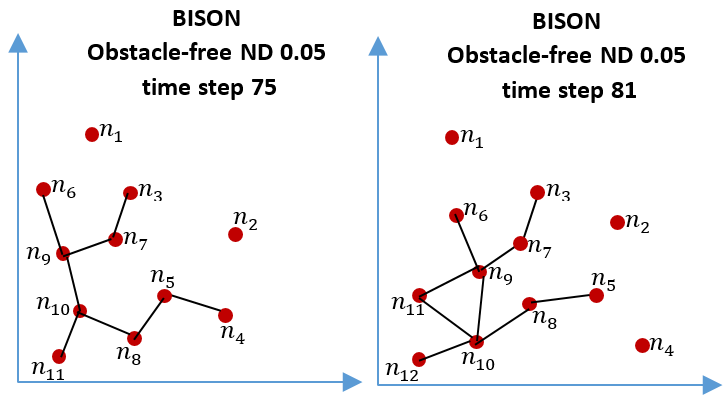}
    \includegraphics[width = \hsize]{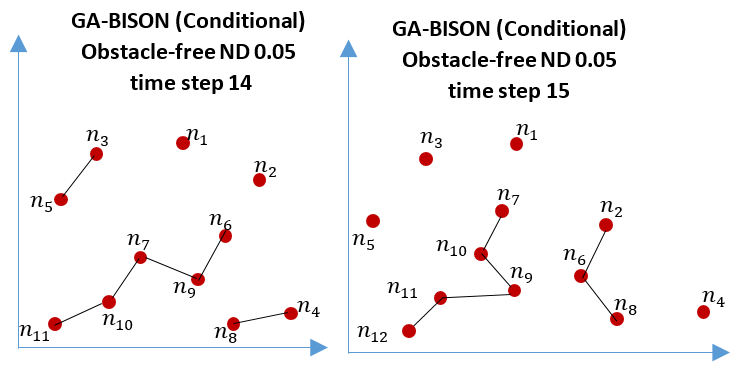}
    \caption{Snapshots of a network generated by (upper) the Voronoi-only algorithm in an obstacle-free environment, at different time steps and (lower) the GA-BISON algorithm in an obstacle-free environment at comparable time steps. ND signifies noise level (among options \ensuremath{0.01, 0.05, 0.1}). See Eq. \ref{eq:PofX}.} 
    \label{fig:NetworkGeneration} 
\end{figure}

\pagebreak
The proposed \textit{GA-BISON (Conditional)} approach allowed each sensor node to decide whether to apply GA or to stay reliant on BISON to determine its next position by checking its number of neighbors. If the sensor has between 1 and 3 neighbors, as shown in Figure \ref{fig:NetworkGeneration} (lower) for node \ensuremath{n_1}, then there is still a chance to move further and discover more regions by implementing a GA approach.

If the sensor node has more than 4 neighbours as in the Figure \ref{fig:NetworkGeneration} for node 9 (upper right panel) there is limited region to discover, unlike node 9 in the lower right panel.
Therefore, it is better for the sensor to optimize its current location by moving towards the Voronoi centroid using BISON. This GA + Voronoi approach improved the execution time and discovery rate of the network, by discovering further locations instead of Voronoi centroids implemented in BISON, but with the price of moderate energy expenditure compared to the Voronoi-only approach. We also validated that GA-BISON (Conditional)'s coverage performance is robust against the effects of noise. Its performance is enhanced by noise but incurs an increased energy cost. From these analyses, the efficiency that we can guarantee from the developed approaches can cover several applications depending on their requirements and abilities of WSN deployment.

\section{Review of related work}\label{sec:SotA}
Temporal networks, also called time-ordered networks or dynamic networks, are general network structures containing timing information about the actions of and interactions between network members \cite{holme2013temporal}. This added information can make analysis significantly more difficult, however the added complexity makes temporal networks natural choices for studying dynamic systems \cite{article31, doi:10.1126/science.aai7488}. 

Temporal network analysis has been used to study a wide variety of topics, including anomalies in urban transportation, disease transmission, and video super-resolution \cite{article33, article34, article35}.
Temporal Network Graphs (TNGs) can be constructed in several different ways. For our purposes here, it is convenient to think of them as a sequence of unweighted, static graphs, each with an accompanying adjacency matrix, defined in a standard manner for static networks, where each element can be written:
\begin{equation}
a_{ij}(t) =
\begin{cases}
    1,  & \text{if node } i \text{ is connected to } j \text{ at time } t \\
    0,  & \text{otherwise}
\end{cases}
\label{eq:AfromItoJ}
\end{equation}

The TNG representation is powerful, as it allows us to utilize familiar static network theoretic techniques on graphs within the sequence, specifically, for each step of the sequence in time evolution of the network. Understanding the interplay between the underlying dynamics and the temporal sampling rate is critical to most temporal network analysis \cite{article31, doi:10.1126/science.aai7488, article36}. Both the Voronoi-only approach and the GA + Voronoi operate at discrete time steps; giving us a natural starting point for sampling times. We then construct TNGs for all trials as the sequence of static graphs occurring at each time step.

Temporal network theory has been explored as a means of identifying shortest time-respecting paths between nodes, making it useful in the development of routing protocols for WSNs \cite{article08, 6514207, article10}. It also provides statistical analysis on the average number of hops and members required from source to destination. Temporal network analysis can also reveal members that are involved in shortest paths, thus playing an important role in mediating the data flow in the network \cite{https://doi.org/10.1111/1365-2656.12764}.
Moreover, temporal and centrality measures are combined in static WSN to decide their best placement based on their response to changes in the surrounding environment \cite{https://doi.org/10.1111/ecog.04716, 6844724}.

Regarding TNG measures and statistics, we consider two broad categories: (a) static measures recorded for individual graphs within the time-ordered sequence, referred to here as time traces, and (b) temporal measures, constructed by looking at a TNG as a single object. Given the added complexity of TNGs, as well as the relative newness of the field, it has been difficult to establish a consensus around the use and definition of measures of type (b) \cite{article31, article36}. We chose to focus a bit more on measures of type (a), largely for the reasons stated above, but recognize that future work utilizing measures of type (b) has significant merit. This being said, we did extract the distribution function of the length connections from the TNGs in the case of our simulated physical situations. Overall, this type of work may also include utilizing aggregate graphs
\cite{article44, Kivelae2018}, supra-centrality measures \cite{article41, ALMUGAHWI2021126121}, and phase transition analysis \cite{article40}.

Static centrality measures are commonly used network measures for categorizing and ranking nodes within a static network, and there exists a large variety of possible measures that can be employed \cite{10.5555/2181136}.

In WSNs, various centralities have been introduced to measure connectivity, clustering/localization, data flow and energy expenditure \cite{7916530, article08, 6514207, article10, SenthilKumaran2021, electronics9050738}. Still other centrality measures are used to show the robustness of the network against errors, particularly in the face of node failure \cite{BORGATTI2006124}. Choosing an applicable centrality measure largely depends on the specific network being studied, its goals, and the environment it exists in.

Among the most frequent centrality measures discussed in literature is the betweenness centrality, a simple measure that identifies nodes working as important corridors for information flow within in the network
\cite{7916530, article14, article08, 6514207, article10, BORGATTI2006124, electronics9050738}. This identification allows for increased control over this information flow, as well as increased reinforcement of important nodes. The limitations of this measure lie in its high computational cost, disregard for the global structure of the network, and its inapplicability to choose the shortest path for data transmission due to limited energy resources
\cite{1624337}.

Another commonly employed centrality measure is the closeness centrality, which measures the mean distance between a node and other nodes (also referred to as the shortest geodesic distance from a node to all other nodes in the network). Closeness centrality is useful in clustering and assigning energy-saving sleep/wake schedules \cite{7916530, article08, 6514207, article10, BORGATTI2006124, SenthilKumaran2021, electronics9050738}. However, it does not account for unreachable nodes and performs poorly in large networks, where it cannot point to the main leaders \cite{1624337}.

We considered 
degree centrality
\cite{https://doi.org/10.1111/1365-2656.12659, JACOBY2016301, 6514207, BORGATTI2006124, SenthilKumaran2021, electronics9050738} and eigenvector centrality \cite{6514207, electronics9050738, BORGATTI2006124, DBLP:journals/corr/OrmanLN17} to be used for WSNs when implemented in toxic-leaks detection as described by Voronoi-type approaches \cite{6514207}. These centralities cover the two types of classifications: (i) local centrality which is demonstrated through degree centrality by focusing on how nodes are connected to their neighbors, and (ii) the global centrality demonstrated through the eigenvector centrality that reflects how often can a node be effective in transferring the data packets among the network \cite{6514207}. Moreover, in \cite{BORGATTI2006124} the analysis of 4 different centrality measures (Degree, Betweenness, Closeness, and Eigenvector) shows them identical in their analysis of robustness against errors in the measurements (addition or removal of nodes) and so choosing either one is adequate for network analysis. Additional research has shown that eigenvector centrality performs better than other centralities at measuring causal inference \cite{Dablander2019}. Complementing all of this is its relative ease of computation.



\section{Bison as a temporal network}\label{sec:BasTN}
Given the changing nature of BISON and its proposed target environments, we believe temporal analysis is necessary to uncover the deeper structural changes within the network in different environmental conditions \cite{https://doi.org/10.1111/1365-2656.12764, article31}. 

For our study, we consider four different environments (cf. Table \ref{tab:table1}). Here, \textit{``moderate''} noise is a \textit{``noise deviation''} level of \ensuremath{0.05} within a simple Gaussian noise model:

\begin{equation}
P(x) = \left(\frac{1}{\sigma \sqrt{2 \pi}}\right) e^{\frac{(x-\mu)^{2}}{2 \sigma^{2}}}
\label{eq:PofX}
\end{equation}

Where \ensuremath{x} is taken to be a random variable, \ensuremath{\mu} is the mean, and \ensuremath{\sigma} is the standard deviation. Increasing \ensuremath{\sigma^{2}} (given by \ensuremath{\sigma^{2} = \frac{N_{o}}{2}}, with \ensuremath{N_{o}} the \textit{noise power}) thus detrimentally impacts signal accuracy. \textit{``Scattering''} obstacles refers to small obstacles placed throughout the environment (see Figure \ref{fig:NetworkGeneration} for the layout). We also limit each algorithm to \ensuremath{40} nodes, both for sake of comparison and to avoid large computation times. It is straightforward to extend this study to \ensuremath{100s} of nodes and to further complexify the environmental conditions, such as number of obstacles, wall penetrability, and the level of noise.

\begin{table}[h]
\caption{\label{tab:table1}%
We analyze each combination of the following parameters, giving us eight different cases to study.}
\begin{ruledtabular}
\begin{tabular}{ccc}
\multicolumn{1}{c}{Algorithm}   &\multicolumn{1}{c}{Noise-level}    &\multicolumn{1}{c}{Obstacle arrangement}\\
\colrule
BISON       &No Noise                                   &No obstacles\\
GA-BISON    &ND \ensuremath{= 0.05}   &10 scatterers\\
\end{tabular}
\end{ruledtabular}
\end{table}

\begin{figure*}[ht]
    \centering
    \includegraphics[width = 0.82\hsize]{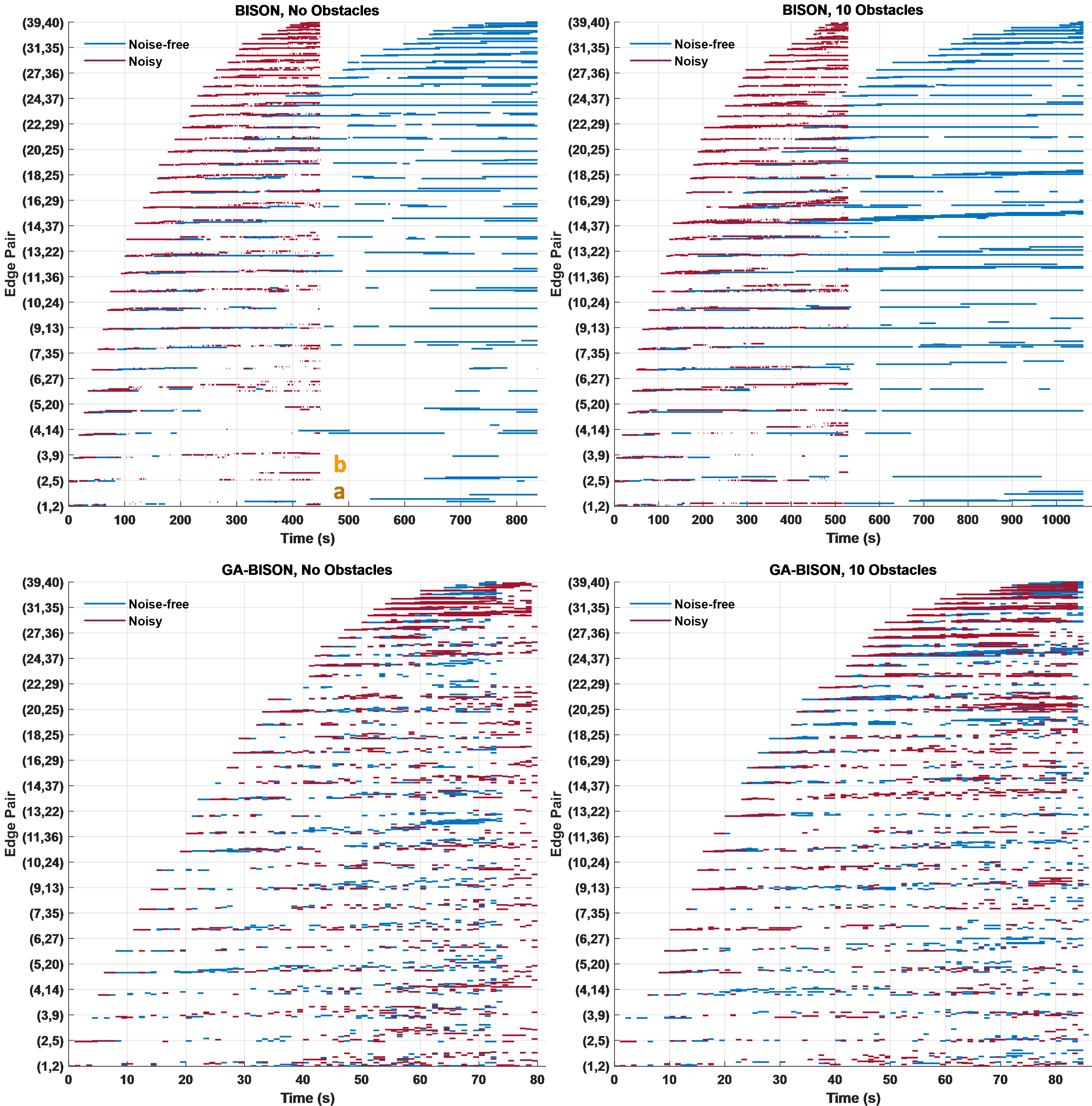}
    \caption{Edge-centric representation of Voronoi-only (BISON) and GA + Voronoi (GA-BISON) in both the obstacle-free environment and obstacle-rich environments. A given simulation has reached the cutoff condition at its rightmost edge. Notice that GA + Voronoi reaches cutoff after a fraction of the time required for Voronoi-only, and how Voronoi-only performs much more quickly in the presence of noise. Also note the sporadic nature of connections within GA + Voronoi, compared to the stability of Voronoi-only connections. \textbf{a} and \textbf{b} mark two of the significant empty bands, where connections never form. Table \ref{tab:table2} lists the connections in \textbf{a} and \textbf{b}, we refer to SOM - C for a complete list of all  missing pairs, for all panels of this figure.}
    \label{fig:EdgeCentricRepresentation} 
\end{figure*}

There are several ways of visually representing the information contained within a TNG. A common representation is the edge-centric representation, which lists the possible edge pairs belonging to the network along the vertical axis, and the temporal dimension across the horizontal axis \cite{holme2013temporal}. When an edge is active, a horizontal line is drawn from the first moment it is active to the last. We generated such temporal network graphs in Figure \ref{fig:EdgeCentricRepresentation}, for varied conditions of numerical experiments, in order to visualize the connectivity actions influenced by various obstacles and noise variations.

In both the scatterer and the scatterer-free case, Voronoi-only reaches the node cutoff much quicker when there is noise present. It is also immediately apparent that the noise-free cases contain persistent, longer connections, while the noisy cases are more sporadic. This graph representation is quite sparse in the vertical dimension, as can be seen from the existence of relatively wide \textit{``empty bands''} of nodes whose connections never form. This sparseness can be attributed to the unlikelihood of nodes encountering nodes from a different time step of the simulation. That is, node \ensuremath{1} is unlikely to encounter node \ensuremath{40}, owing to the large amount of distance traversed by node \ensuremath{1} before node \ensuremath{40} is inserted. 

\pagebreak
\begin{table}[b]
\caption{\label{tab:table2}%
: Edge pairs located within gaps \textbf{a} and \textbf{b} in Figure \ref{fig:EdgeCentricRepresentation}. These gaps indicate edges that never form. SOM - C contains all the missing pairs, for all four panels of this figure.} 
\begin{ruledtabular}
\begin{tabular}{ll}
\multicolumn{1}{c}{\textbf{a}}   &\multicolumn{1}{c}{\textbf{b}}\\
\colrule
(1,4)                               &(2,5), (2,6), (2, 7)\\
(1,6), (1,7), (1, 8)                      &(2,9)\\
(1,10), (1,11), (1, 12)                    &(2,11), (2,12), \ensuremath{\ldots}, (2, 40)\\
(1,14), (1,15), \ensuremath{\ldots}, (1, 19)  &\\
(1,21), (1,22), \ensuremath{\ldots}, (1, 40)  &\\
\end{tabular}
\end{ruledtabular}
\end{table}

~\\

It is worth noting that this 
brings up a representation problem in Figure \ref{fig:EdgeCentricRepresentation}: the graphs are significantly sparser at the bottom, not necessarily because of lower clustering, but rather because there are more edges that start with low numbers 
(no edge is double counted).

The networks resulting from GA + Voronoi are much more sporadic than any of those seen in the Voronoi-only cases, a behavior consistent across various environmental conditions.  We see a similar vertical sparseness.

It is also worth noting that the required number of steps for node cutoff is significantly less than any of the Voronoi-only runs; this is the case regardless of noise.

\begin{figure*}[t]
    \centering
    \includegraphics[width = 0.78\hsize]{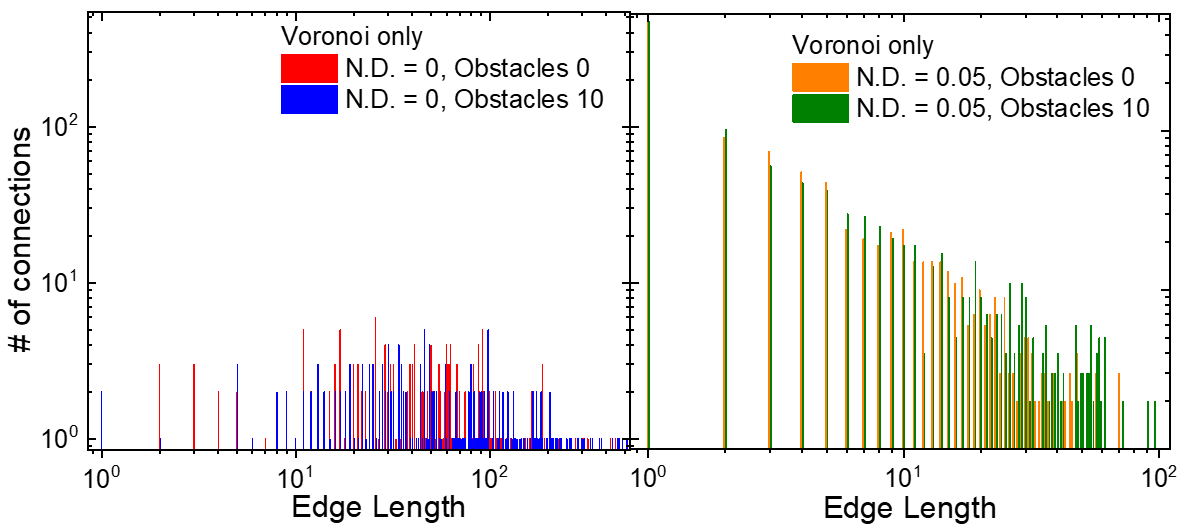}
    \includegraphics[width = 0.78\hsize]{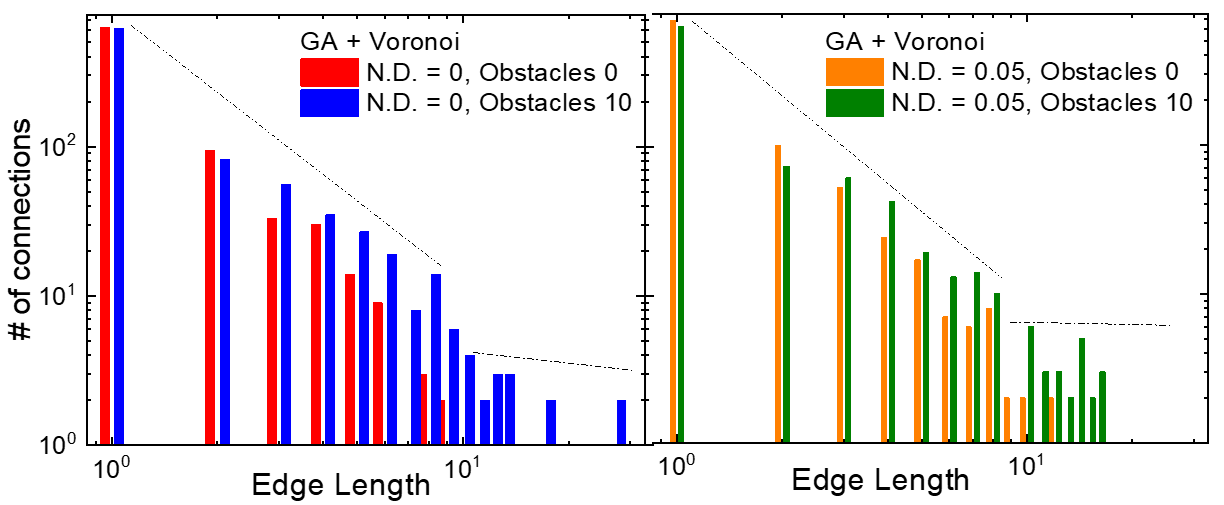}
    \caption{Distribution of various lengths of connection for each of the cases in Figure \ref{fig:EdgeCentricRepresentation}. We see a drastic qualitative difference between Voronoi-only in noise-free and noisy environments, and smaller quantitative difference between GA + Voronoi cases.}
    \label{fig:DistributionOfLength} 
\end{figure*}

The comparison between different environments using Voronoi-only showed that noise affects the simulation time steps required to achieve full coverage in the network and affects the connectivity between the nodes together. In other words, noise allowed nodes to be more distributed in the region, however, created some disconnections between the nodes throughout the time steps, affecting the flow of data packets between the nodes.

Among (almost) any network practical considerations are parameters like (i) energy needed to operate and maintain network connections, (ii) the number of the connections and their distribution in time. Such parameters are easily accessible using the representation in Figure \ref{fig:DistributionOfLength}.

The behavior of GA + Voronoi 
showed that the connectivity is fluctuating more often in obstacle-free than in obstacle-rich environments. This is because in obstacle-free environments, sensor nodes are more freely to move and disperse, losing by that their connectivity more easily than in obstacle-rich environments. The temporal network figures showed significant differences in the behavior between Voronoi-only and GA + Voronoi.

The first difference to discuss is the time steps. We can notice that in GA + Voronoi, 
the number of time steps is not affected by noise, however in Voronoi-only, the number of time steps decreases in the presence of noise. This indicates that merging GA with Voronoi introduced a robustness functionality to Voronoi-only against communication noise, at least in terms of the execution time.

The second difference is regarding the number of connections established in the network. We observed that the presence of noise in the environment allowed more connections to be present between the sensor nodes compared to noise-free environment. This behavior is observed in both approaches (Voronoi and GA + Voronoi) with a slightly more connections available in Voronoi-only compared to GA + Voronoi 
approach. 

The third noticed difference is the connectivity duration among the agents in the network. For GA + Voronoi, the system suffered from un-stabled connections throughout the simulation, but with less effect in the presence of noise than in noise-free situations. In Voronoi-only we observed the opposite behavior, where in noise-free condition, the connections are established at earlier stages of the simulation and are more stabled; while in noise-rich condition, un-stabled connections are performed throughout the simulation which stabilize as the number of nodes increase in the network over time.

\begin{figure*}[t]
    \centering
    \includegraphics[width = 0.7\hsize]{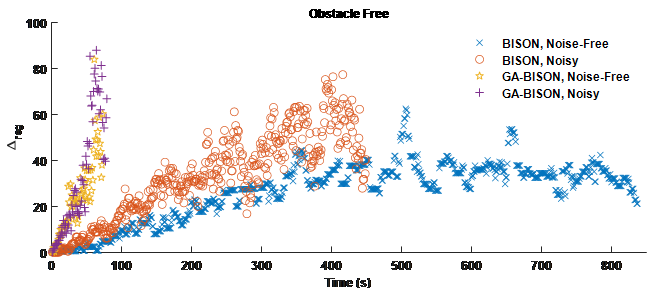}
    \includegraphics[width = 0.7\hsize]{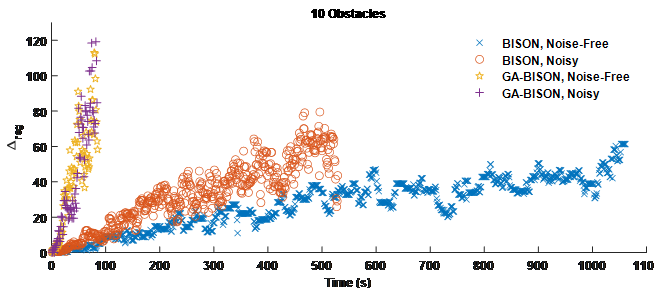}
    \caption{Temporal evolution of the deviation from regularity for Voronoi-only and GA + Voronoi approaches under different environmental conditions. The measure \ensuremath{\Delta_{\text{reg}}} is defined in Eq. \ref{eq:DeltaReg}, below.}
    \label{fig:TemporalEvolution} 
\end{figure*}
\section{Regularity measure}\label{sec:measureRegularity}

Further observations of Voronoi-only and GA + Voronoi network-time evolution motivates an elementary inquiry into the regularity of networks deployed. 

A network is considered regular if
\cite{10.5555/2181136}:

\begin{equation}
\sum_{i=1}^{n} \lambda_{i}^{2} = n\lambda_{\text{max}}
\label{eq:SigmaLambda}
\end{equation}
where \ensuremath{\lambda_{i}} are the eigenvalues of the adjacency matrix, \ensuremath{\lambda_{\text{max}}} is the largest eigenvalue, and \ensuremath{n} is the number of nodes in the network, all of which are recorded at a particular time step. 

We define the \textit{regularity difference} at a given time as:
\begin{equation}
\Delta_{\text{reg}} = \displaystyle \left\lvert \sum_{i=1}^{n} \lambda_{i}^{2} - n\lambda_{\text{max}}\right\rvert
\label{eq:DeltaReg}
\end{equation}

Figure \ref{fig:TemporalEvolution} represents the results of the network regularity at different moments in time and for several situations (obstructed environments and noise levels). 

We can notice that for both approaches at earlier steps of the network evolution, the network is closer to regularity than at further time steps. 

Moreover, by looking at the moments of injecting new nodes, the regularity difference (\ensuremath{\Delta}) increases in the network in most environments. Regularity difference levels are dramatically different between the GA + Voronoi and Voronoi-only cases. 
Furthermore, noise seemingly increases the regularity difference (\ensuremath{\Delta}) in Vornoi-only networks, while the GA + Voronoi networks maintain similar regularity difference levels in the face of noise.

Below, we will discuss how the sharply different network irregularities correlate with the practical performance measures, as well as energy expenditure.


\section{Centrality measure}\label{sec:measureCentrality}

The motivation of this work is driven by the need to understand and improve the network coverage in the context of the spatial, temporal and energy constraints. Specifically, implementations are often driven by:
\begin{enumerate}
\item cumulative distance traveled (CDT) by the nodes
\item percent area coverage (PAC) that nodes manage to cover by the signal
\item minimal time to achieve target coverage
\item optimized energy allowed for the network of drones 
\end{enumerate}

\begin{figure}[t]
    \centering
    \includegraphics[width = \hsize]{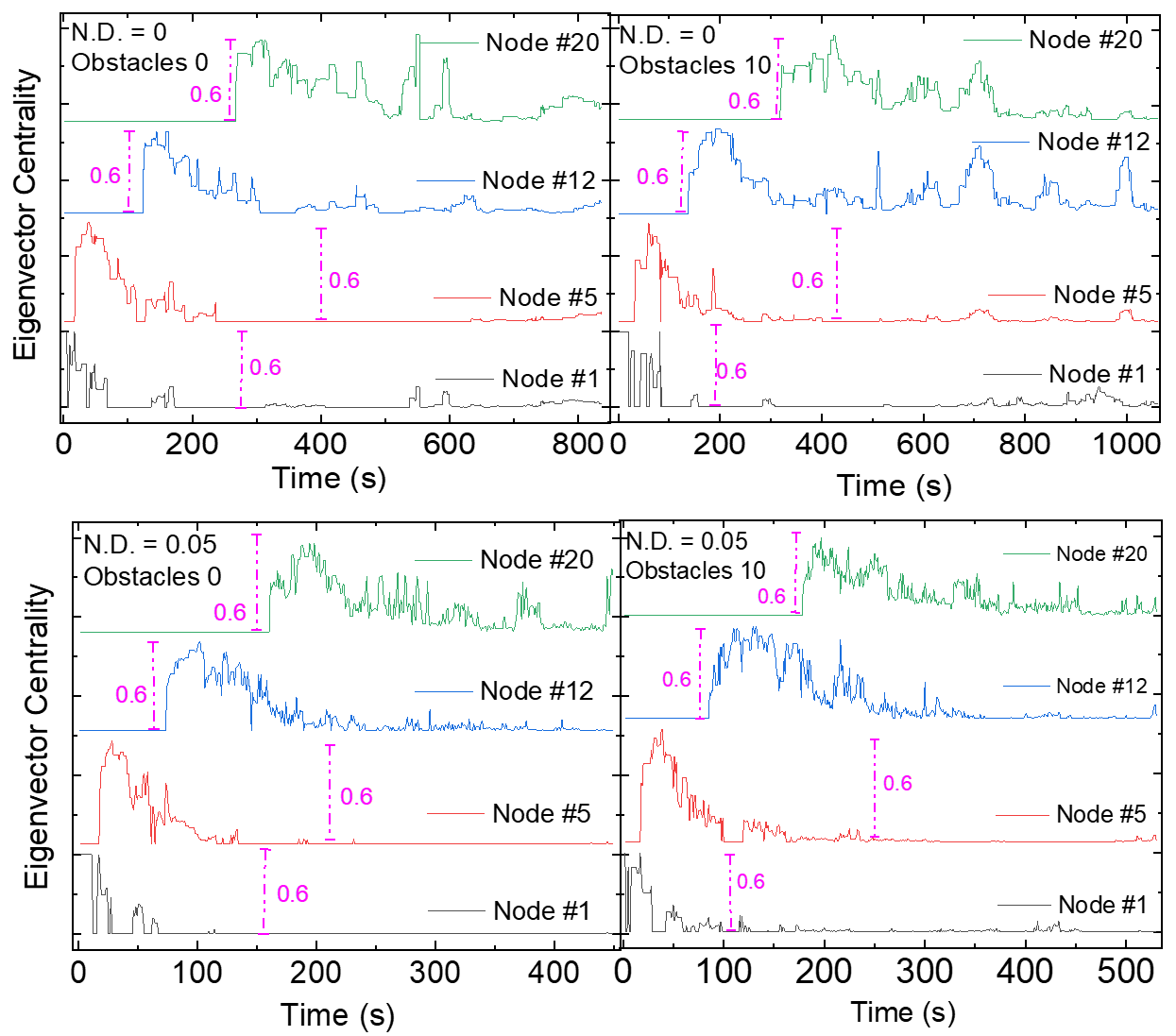}
    \caption{Eigenvector Centralities of the Voronoi-only approach in various environments at different time steps; Vertical bar sign \textit{``}\ensuremath{0.6}\textit{''}, stacked vertically, is common to all curves.}
    \label{fig:EigenVectorCentralityBISON} 
\end{figure}

The analysis so far illustrates how uneven the roles of most nodes are during the deployment process. We also noted that as the deployment progresses, the nodes that have few links can quickly gather many links, indicating that a fair bit of temporal switching occurs within the networks. 
With the goal of initial quantification of the role of individual nodes and the time-dependent role they play, we determined eigenvector centrality (\textit{EC}) \cite{10.5555/2181136}, one of the leading measures that help understands the evolving nature of temporal network graphs. 

We select eigenvector centrality as it is sufficiently general (typically defined as a generalization of Katz index \cite{10.5555/2181136}), applicable to smaller size networks (N \ensuremath{\approx} 30-100 in most our cases), and is readily obtainable from the adjacency matrices. It measures the importance of a node based on the importance of its (connected) neighbors.

Comparing our values with \cite{article08} we notice that local centrality measures (our degree centrality measures and their betweenness centrality measures) provide high spike values for nodes that are more connected to their neighbors than others, while global centralities (our eigenvector centrality measures and their closeness centrality measures) have a smooth distribution of values among the nodes. 
We also observed that the differences in the eigenvector centralities of the nodes in the network at different time steps revealed the uneven role of most of the nodes in the network during their deployment process, which is considered an important finding to be used later to keep track of the important nodes that can be relied on to transfer data packets among the network members.

We present eigenvector centrality measurements for the various BISON cases in Figure \ref{fig:EigenVectorCentralityBISON} for the selected environmental cases. Nodes with early injection time immediately spike in importance, but drop off as deployment continues. By contrast, nodes deployed later spike early as well, and then maintain some non-zero importance throughout their lifetime. This initial surge is to be expected, since nodes are injected from the same corner, resulting in an area with a high node density.

\begin{figure}[t]
    \centering
    \includegraphics[width = \hsize]{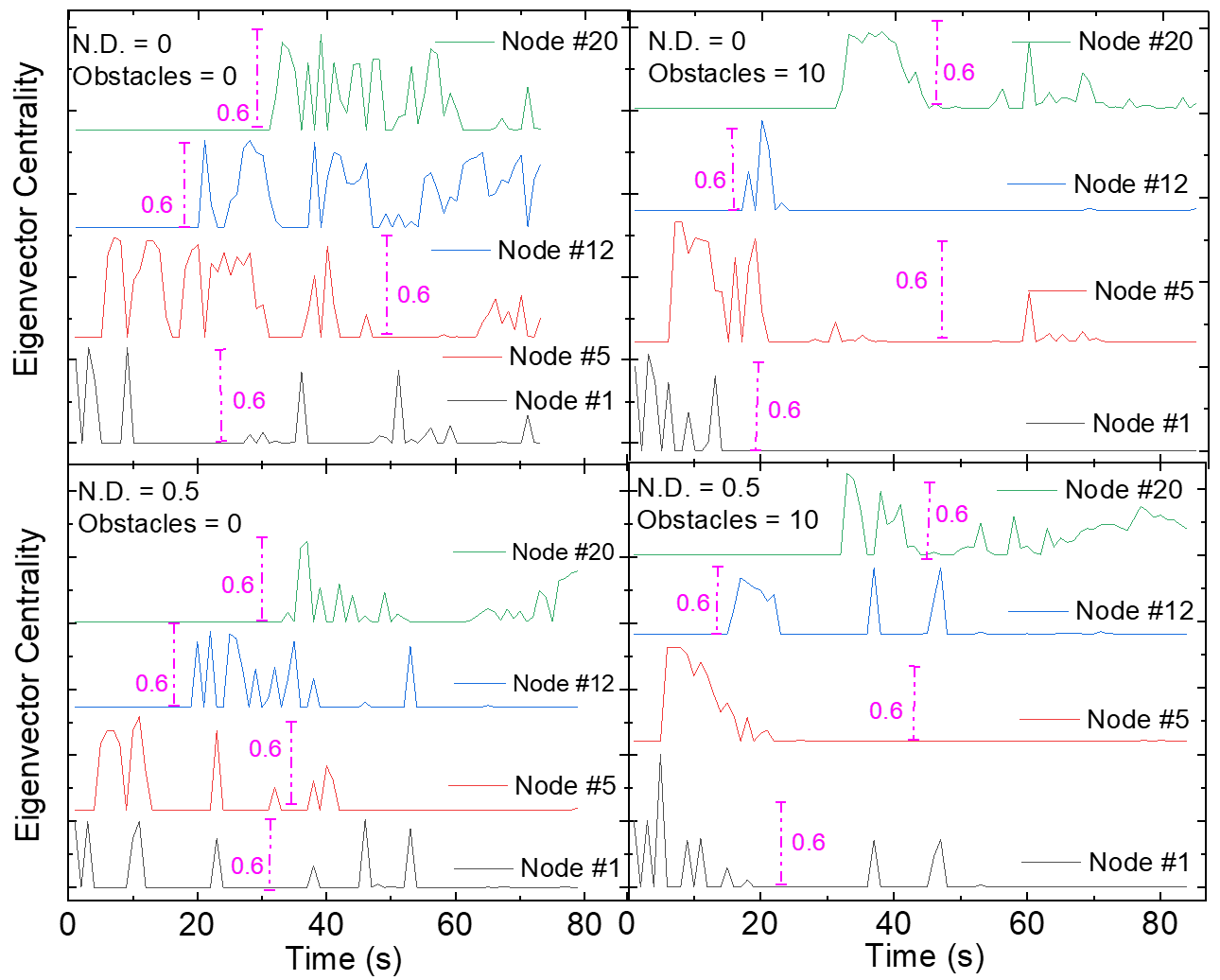}
    \caption{Eigenvector Centralities of GA + Voronoi in various environments at different time steps; Vertical bar sign \textit{``}\ensuremath{0.6}\textit{''} is common to all curves, which are stacked vertically for clarity.}
    \label{fig:EigenVectorCentralityGA-BISON} 
\end{figure}

As spreading continues, early nodes migrate to distant regions, and thus have little opportunity to be deeply connected, while later nodes remain relatively close to this highly dense region. These EC findings also clearly support our previous insight that adding noise does, in fact, have a beneficial impact on the performance of Voronoi-only generated networks with regard to deployment speed. When comparing plots from simulations differing only in noise level, we consistently see that the presence of noise smooths the EC time trace curve, suggesting a more homogeneous distribution of the nodes. We see somewhat similar spiking behavior in the GA + Voronoi trials featured in
Figure \ref{fig:EigenVectorCentralityGA-BISON}, although early deployment nodes are more likely to feature significant spikes in EC later in their lifetimes. This is partly expected, due to the somewhat randomized nature of the GA part of the algorithm.

In neuroscience, EC has been found to correlate with a neuron's firing rate and in cellular biophysics with \cite{sneppen_2014}. We conjecture an analogous phenomenon could be exploited here;  there may be an advantage in designing our WSN nodes such that their message passing rate is highest immediately after their injection, when a node can be expected to have high EC. After some time, this transfer rate can drop to some consistent, energy-efficient base value. The potential benefit that can be gained by this should be measurably higher for environments where there is noise, since in such environments there are consistently fewer - and lower – \textit{``flare-ups''} of EC values as time progresses. 

We have analyzed correlations between the eigenvector centrality time traces from Figure \ref{fig:EigenVectorCentralityGA-BISON}, using the well known formula for Pearson's correlation coefficient, where we have taken the expectation values of our data to be the arithmetic mean of our data sets. For two data sets \ensuremath{A} and \ensuremath{B}, the coefficient is

\begin{equation}
C_{A,B} = \frac{\langle AB \rangle - \langle A \rangle\langle B \rangle}
        {\sqrt{\langle A \rangle^{2} - \langle A^{2} \rangle} \sqrt{\langle B \rangle^{2} - \langle B^{2} \rangle}}
\label{eq:Pearson}
\end{equation}

\begin{figure}[h]
    \centering
    \includegraphics[width = \hsize]{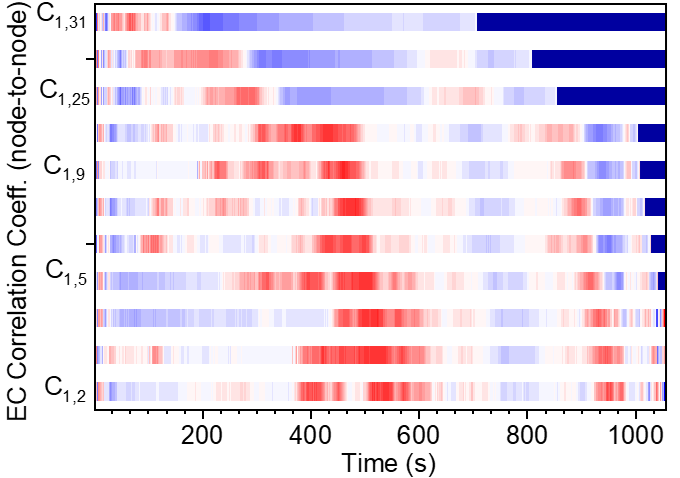}
    \includegraphics[width = \hsize]{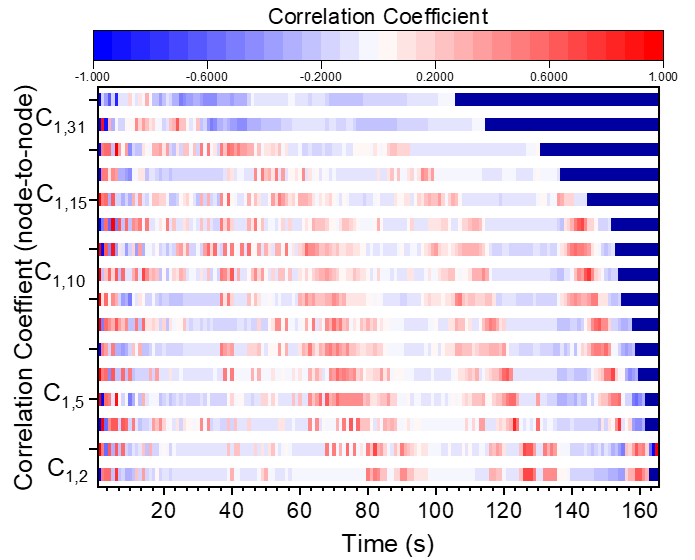}
    \caption{Time-lag correlation heat maps for time-dependent eigenvector centralities from Figures \ref{fig:EigenVectorCentralityBISON} and \ref{fig:EigenVectorCentralityGA-BISON}. (selected node pairs shown only – more data in SOM). (upper) Voronoi; 
    (lower) GA + Voronoi; (both) ND \ensuremath{= 0.05} and 10 obstacles.}
    \label{fig:TimeLagCorrelation} 
\end{figure}

The outcome of that analysis is shown in Figure \ref{fig:TimeLagCorrelation}. 

Having seen the overall shape and detailed behavior for a number of eigenvector centralities, we decided to examine the time-lag correlations of the time traces of eigenvector centralities for selected nodes. We see this as an alternative method of presenting and analyzing node ranking. If one imagines this work as implemented on board a number of (micro-)drones to make their navigation and time-to-task efficient, having several approaches to assessing time-evolving network functionality. 

Keeping in mind that one way to think about eigenvector centrality is that it serves as method for ranking nodes, it is understandable that (once enough nodes are present in the network) increase in the ranking of node \ensuremath{k}, there is a decrease in the ranking for one or more nodes \ensuremath{k-p}, \ensuremath{k+p}. Positive correlation seems to arise when the nodes are close to becoming connected, are connected, or have been connected, where negative correlation seems to suggest the opposite. 

Since the correlation itself depends on time, one can learn quite a bit by reading out the map. It is clear there are \textit{``correlation bands''} (of steadily positive or negative correlation), often  preceded by rapid switching between positive and negative correlation. In the Figure \ref{fig:TimeLagCorrelation} we have omitted quite a few node-to-node EC correlations, for the sake of clarity. Supporting Online Materials contain more examples, omitted from the main text in the interest of space. Figure \ref{fig:TimeLagCorrelation} contains \textit{``conditions rich''} examples, 
with both, noise and obstacles, for reaslism.



\section{Conclusion}\label{sec:Conclusion}
We demonstrated how to apply quantitative methods of temporal network analysis to our Voronoi-like algorithms in order to compare and analyze the dynamics of WSN coverage and robustness in the face of environmental noise and obstructions. GA + Voronoi in particular demonstrated robustness in the face of thee various environmental conditions. To summarize, we can provide the following answers to our research questions in section \ref{sec:Intro}:
\begin{enumerate}
\item With Voronoi-only, nodes move outwards during their deployment in a regular fashion, and their importance to connectivity and information relaying drops off with this outward movement. GA + Voronoi demonstrates similar behavior; however, nodes are more likely to reestablish their importance later.
\item We see that Voronoi-only in a noise-free environment maintains a small number of long-lasting connections. Adding noise substantially redistributes those connection lengths. By contrast, GA + Voronoi maintains qualitatively similar connection length distributions through a variety of environments, featuring a rapidly changing topology with short, frequent connections between nodes. In all cases, nodes are highly likely to associate only with nodes sharing a similar deployment time
\item We see that noise increases the deviation from regularity, smooths out the EC time trace, and impacts the connection length distribution in the Voronoi-only cases. By contrast, noise has a much smaller impact on the same measures applied to the GA + Voronoi cases. This supports previous work that had indicated Voronoi-only changes substantially in the face of noise, while GA + Voronoi is robust to such environmental changes.
\item Utilizing temporal network characteristics allows us to measure and observe the behavior of the network as a whole and if individual nodes.
\item All this was possible without the explicit use of physical measurements of the nodes’ coverage or their environment. These measures also suggest some kind of deeper functional equivalence between Voronoi-only in noisy environments and GA + Voronoi.
\end{enumerate}

Such quantification measurements may prove useful in further refining our algorithms, by allowing us to fine-tune communication routing protocols, broadcasting strength, and other implementation details based on the expected behavior.

While mapping these theoretical network measures onto applied performance metrics can perhaps be challenging, we do believe there are some clear points to take note of. When discussing regularity, we noted that regularity difference rose in the noisy Voronoi-only cases and GA + Voronoi cases. This irregularity seems to have some correspondence to speed of coverage, and we posit that such highly irregular structures may be crucial in this rapid deployment. By contrast, the lower regularity difference of Voronoi-only likely corresponds to its slower but more consistent nature. This parallels the behavior we see in measuring percent area covered (PAC), with GA + Voronoi rapidly reaching high levels of PAC, while Voronoi-only slowly converges towards an acceptable value.

We also discussed EC among nodes in the various cases. By looking at EC, we see that nodes consistently move towards positions of lower importance, especially in the Voronoi-only cases. Fluctuations following this are indicative of a node relocating/reorganizing itself in the network and are much more common in the application of GA + Voronoi.  This fluctuating corresponds to the cumulative distance traveled (CDT) measures. Assuming a fluctuating EC value corresponds to a particularly mobile node, we would expect the distance travel to be much higher for cases with fluctuating EC. Indeed, our previous research indicated GA + Voronoi nodes travel significantly longer distances, seemingly matching this explanation.

In addition to time evolution of irregularity (Section \ref{sec:measureRegularity}) and  time evolution of eigenvector centrality (Section \ref{sec:measureCentrality}), we analyzed the nodes degree distribution, where analogous observations  could be made. In the interest of brevity, this output is presented in SOM D, E.


\begin{acknowledgments}
We acknowledge early support for this project under the UAE ICTFund grant ``Bio-inspired Selforganizing Networks''. KE is thankful for the Graduate Students Fellowship at KU. A part of the work was performed while AFI was a visiting scientist at Cornell University, the hospitality of which is greatly acknowledged. NDB acknowledges the Colgate Undergraduate Research Fund. We acknowledge early stimulating conversations with Dr. D. Ruta (KUST, Abu Dhabi, UAE) and Dr. F. Saffre (VTT, Finland).

NDB and KE contributed in equal parts to the work reported in this paper.
\end{acknowledgments}


\bibliographystyle{apsrev4-2}
\bibliography{PhysRev.DiBrita.2021.Bibliography.bib}
\newpage
\widetext
\begin{center}
\textbf{\large Supporting Online Materials (SOM)}
\end{center}
\setcounter{equation}{0}
\setcounter{figure}{0}
\setcounter{table}{0}
\setcounter{page}{1}
\makeatletter
\renewcommand{\theequation}{S\arabic{equation}}
\renewcommand{\thefigure}{S\arabic{figure}}
\renewcommand{\bibnumfmt}[1]{[S#1]}
\renewcommand{\citenumfont}[1]{S#1}

\section*{SOM – A: A Part of Voronoi Tessellation Process and Project Outline}
\subsection*{Illustrative description of the evolution of nodes in the unknown region}
\begin{figure}[h!]
    \centering
    \includegraphics[width = 0.9\hsize]{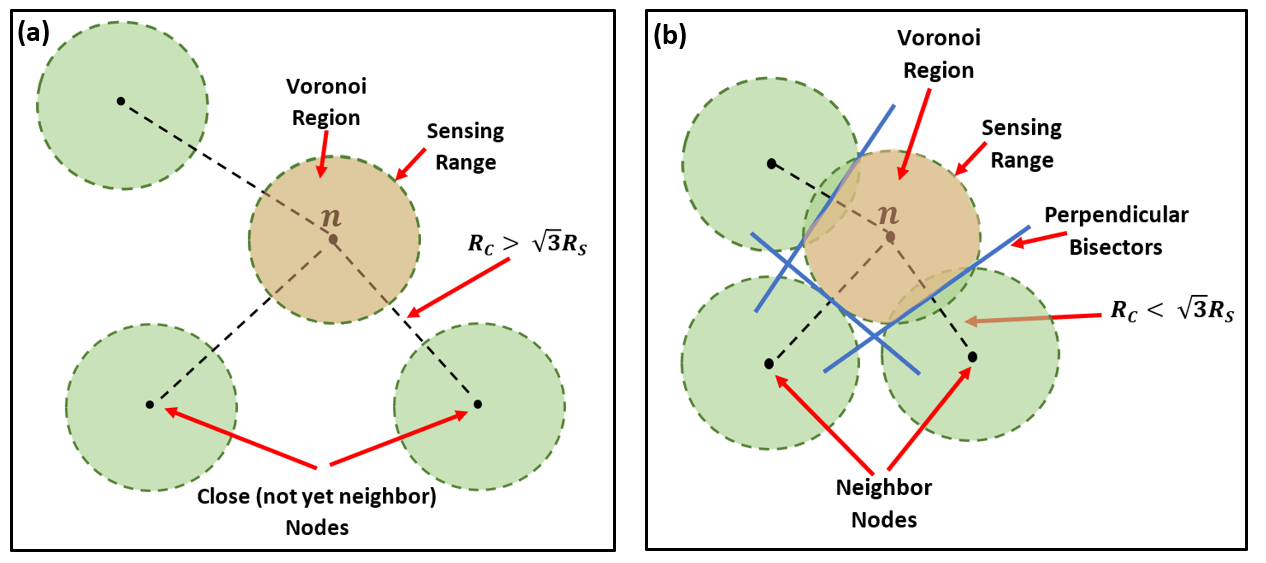}
\end{figure}

Illustrative description of the evolution of nodes in the unknown region, and the generation of the Voronoi region for the next deployment step using our version of Voronoi tessellation - BISON. (a) illustrates the disconnected stage of the nodes, where the Voronoi region is the same as the sensing region. (b) nodes establish perpendicular bisectors to identify their Voronoi region and their next position. \ensuremath{R_{s}} is the sensing range of the moving sensors, while \ensuremath{R_{c}} the center to center distance between two sensors.

\newpage
\subsection*{Organization of the paper}
\begin{figure}[h!]
    \centering
    \includegraphics[width = 0.9\hsize]{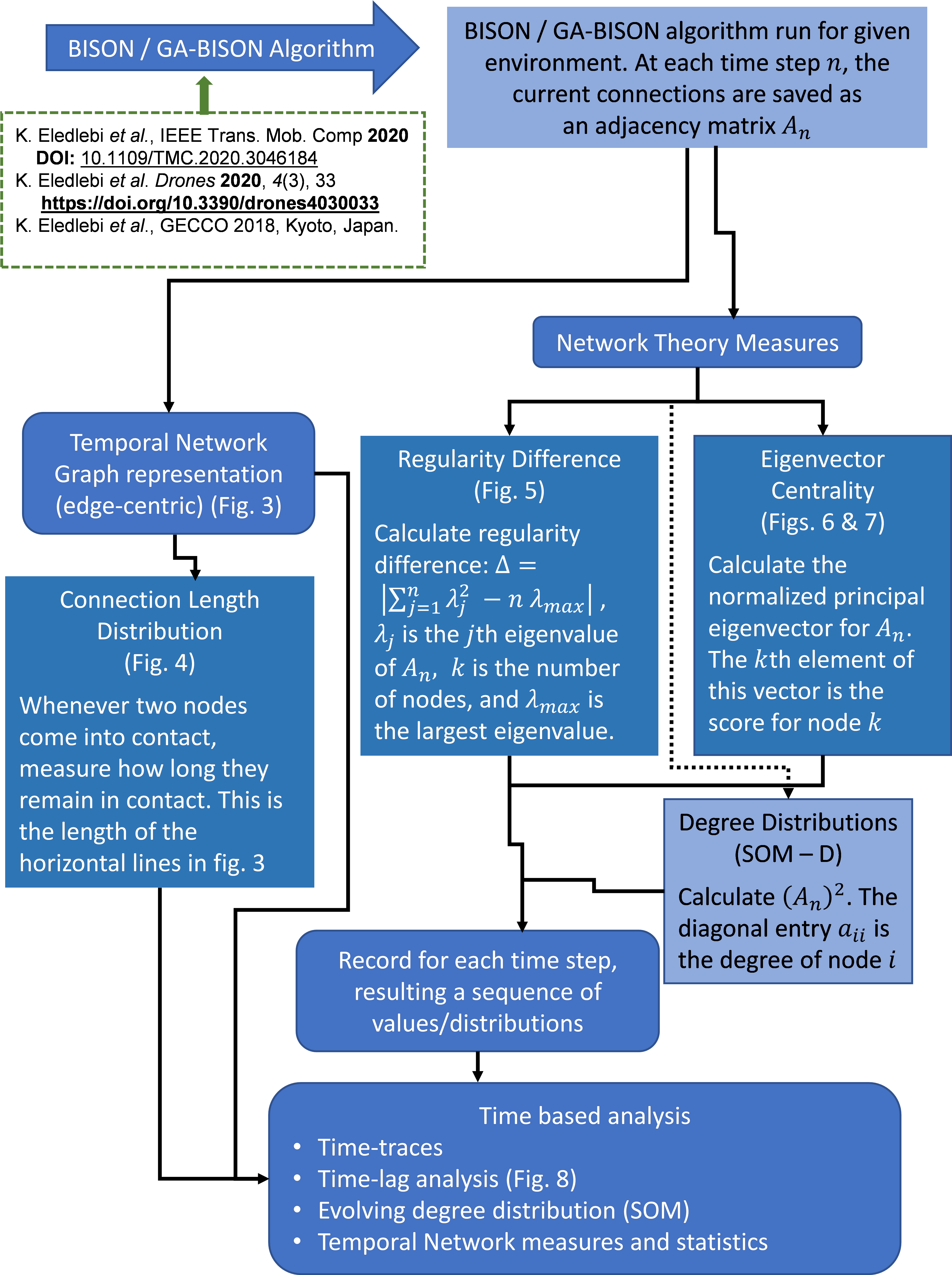}
\end{figure}

\newpage
\section*{SOM B}

\begin{figure}[h!]
    \centering
    \includegraphics[width = \hsize]{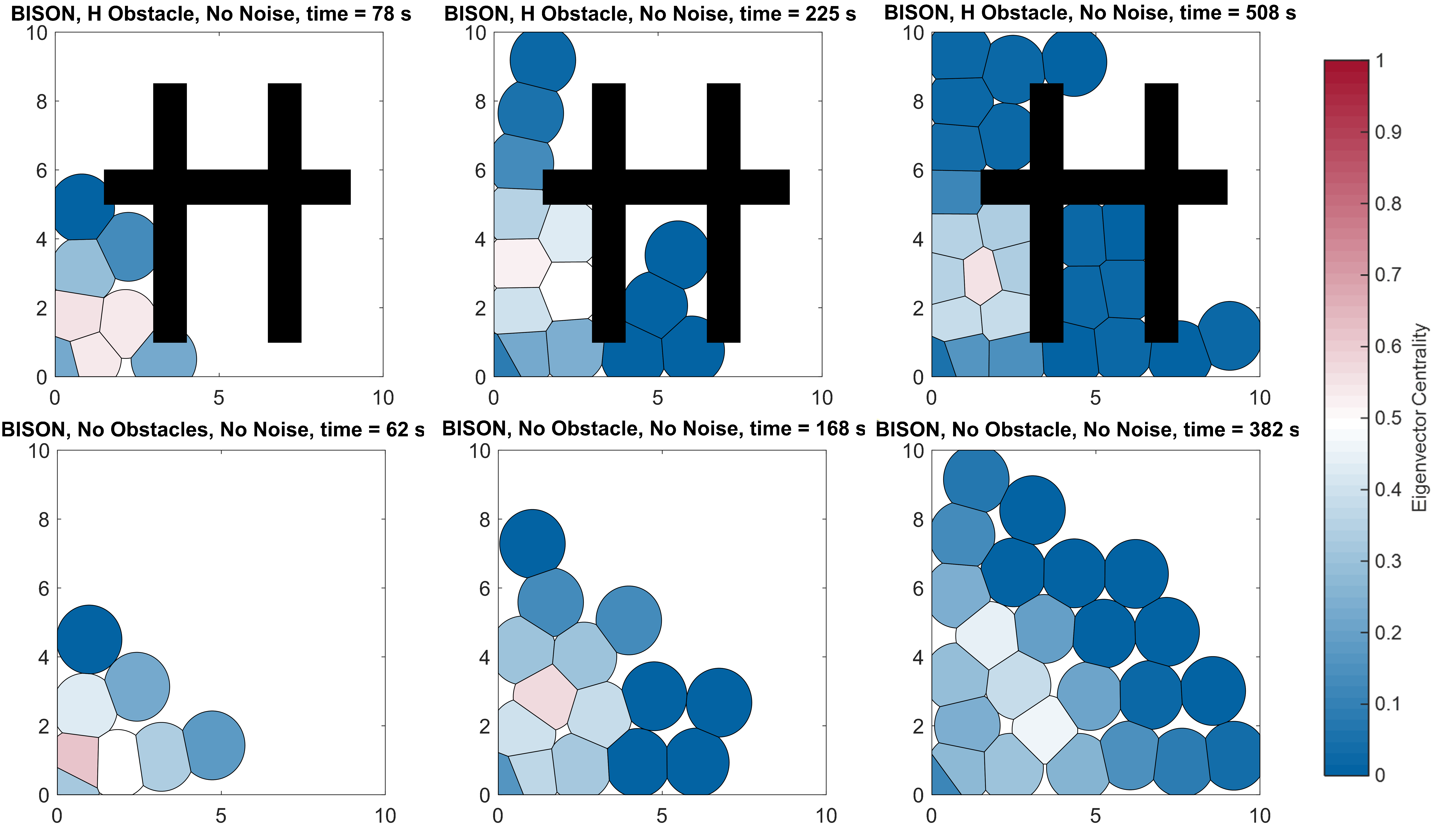}
\end{figure}

BISON (and GA-BISON) is adaptable to other kinds of environments. BISON is shown here exploring regions containing different Obstacles. This figure is related to Fig. 1 in the main text.
\newpage
\section*{SOM C}
\begin{longtable}{|l|l|l|l|}
\hline
\multicolumn{4}{|c|}{BISON}\\
\multicolumn{2}{|c|}{No Obstacles}&\multicolumn{2}{|c|}{10 Scatterers}\\
\multicolumn{1}{|c}{No Noise}&\multicolumn{1}{|c}{Noisy}&\multicolumn{1}{|c}{No Noise}&\multicolumn{1}{|c|}{Noisy}\\
\hline
(1,4)            &(1,6)             &(1,5)&(1,4)\\
(1,6)\ensuremath{\ldots}(1,8)    &(1,7)             &(1,7)\ensuremath{\ldots}(1,9)&(1,5)\\
(1,10)\ensuremath{\ldots}(1,12)  &(1,9)\ensuremath{\ldots}(1,40)    &(1,12)\ensuremath{\ldots}(1,20)&(1,7)\ensuremath{\ldots}(1,40)\\
(1,14)\ensuremath{\ldots}(1,19)  &(2,5)             &(1,22)\ensuremath{\ldots}(1,24)&(2,6)\\
(1,21)\ensuremath{\ldots}(1,40)  &(2,8)\ensuremath{\ldots}(2,16)    &(1,26)\ensuremath{\ldots}(1,40)&(2,7)\\
(2,5)\ensuremath{\ldots}(2,7)    &(2,18)\ensuremath{\ldots}(2,40)   &(2,7)\ensuremath{\ldots}(2,9)&(2,9)\ensuremath{\ldots}(2,17)\\
(2,9)            &(3,6)\ensuremath{\ldots}(3,8)     &(2,12)\ensuremath{\ldots}(2,40)&(2,19)\ensuremath{\ldots}(2,40)\\
(2,11)\ensuremath{\ldots}(2,40)  &(3,10)            &(3,6)&(3,8)\\
(3,6)            &(3,13)\ensuremath{\ldots}(3,40)   &(3,8)\ensuremath{\ldots}(3,40)&(3,10)\ensuremath{\ldots}(3,40)\\
(3,8)\ensuremath{\ldots}(3,40)   &(4,9)\ensuremath{\ldots}(4,13)    &(4,6)&(4,9)\\
(4,7)            &(4,15)\ensuremath{\ldots}(4,40)   &(4,7)&(4,11)\ensuremath{\ldots}(4,17)\\
(4,9)\ensuremath{\ldots}(4,11)   &(5,10)            &(4,9)\ensuremath{\ldots}(4,40)&(4,19)\\
(4,13)\ensuremath{\ldots}(4,20)  &(5,12)\ensuremath{\ldots}(5,14)   &(5,8)&(4,22)\\
(4,22)\ensuremath{\ldots}(4,40)  &(5,16)\ensuremath{\ldots}(5,40)   &(5,10)\ensuremath{\ldots}(5,40)&(4,23)\\
(5,8)            &(6,9)             &(6,9)&(4,25)\ensuremath{\ldots}(4,40)\\
(5,10)           &(6,11)            &(6,11)\ensuremath{\ldots}(6,40)&(5,9)\\
(5,12)           &(6,12)            &(7,10)&(5,11)\ensuremath{\ldots}(5,40)\\
(5,14)\ensuremath{\ldots}(5,40)  &(6,14)\ensuremath{\ldots}(6,16)   &(7,11)&(6,10)\\
(6,9)            &(6,18)\ensuremath{\ldots}(6,21)   &(7,13)\ensuremath{\ldots}(7,16)&(6,12)\\
(6,11)\ensuremath{\ldots}(6,40)  &(6,23)\ensuremath{\ldots}(6,40)   &(7,18)&(6,14)\\
(7,12)\ensuremath{\ldots}(7,40)  &(7,9)             &(7,19)&(6,16)\\
(8,11)           &(7,11)            &(7,21)&(6,18)\\
(8,13)\ensuremath{\ldots}(8,15)  &(7,12)            &(7,23)\ensuremath{\ldots}(7,40)&(6,20)\ensuremath{\ldots}(6,40)\\
(8,17)\ensuremath{\ldots}(8,20)  &(7,14)\ensuremath{\ldots}(7,21)   &(8,12)&(7,10)\\
(8,22)           &(7,23)\ensuremath{\ldots}(7,40)   &(8,14)&(7,14)\ensuremath{\ldots}(7,40)\\
(8,24)\ensuremath{\ldots}(8,40)  &(8,12)            &(8,15)&(8,13)\ensuremath{\ldots}(8,15)\\
(9,12)           &(8,16)\ensuremath{\ldots}(8,40)   &(8,18)&(8,17)\\
(9,14)           &(9,13)\ensuremath{\ldots}(9,17)   &(8,19)&(8,19)\ensuremath{\ldots}(8,40)\\
(9,16)\ensuremath{\ldots}(9,40)  &(9,19)\ensuremath{\ldots}(9,40)   &(8,21)&(9,14)\ensuremath{\ldots}(9,16)\\
(10,13)          &(10,15)\ensuremath{\ldots}(10,40) &(8,23)\ensuremath{\ldots}(8,40)&(9,18)\ensuremath{\ldots}(9,40)\\
(10,15)          &(11,13)           &(9,13)\ensuremath{\ldots}(9,16)&(10,13)\\
(10,17)\ensuremath{\ldots}(10,40)&(11,17)\ensuremath{\ldots}(11,20) &(9,18)&(10,15)\\
(11,15)\ensuremath{\ldots}(11,17)&(11,22)\ensuremath{\ldots}(11,40) &(9,19)&(10,17)\ensuremath{\ldots}(10,40)\\
(11,19)\ensuremath{\ldots}(11,40)&(12,17)           &(9,21)&(11,14)\\
(12,15)          &(12,19)\ensuremath{\ldots}(12,40) &(9,23)\ensuremath{\ldots}(9,27)&(11,17)\\
(12,18)\ensuremath{\ldots}(12,20)&(13,18)           &(9,29)\ensuremath{\ldots}(9,40)&(11,19)\\
(12,22)\ensuremath{\ldots}(12,40)&(13,21)           &(10,12)&(11,20)\\
(13,16)          &(13,23)\ensuremath{\ldots}(13,40) &(10,15)&(11,24)\ensuremath{\ldots}(11,40)\\
(13,17)          &(14,16)\ensuremath{\ldots}(14,18) &(10,17)&(12,19)\\
(13,19)          &(14,20)           &(10,19)&(12,20)\\
(13,21)\ensuremath{\ldots}(13,40)&(14,22)\ensuremath{\ldots}(14,25) &(10,20)&(12,22)\ensuremath{\ldots}(12,40)\\
(14,17)          &(14,27)\ensuremath{\ldots}(14,40) &(10,22)\ensuremath{\ldots}(10,40)&(13,18)\ensuremath{\ldots}(13,40)\\
(14,20)\ensuremath{\ldots}(14,40)&(15,20)           &(11,15)\ensuremath{\ldots}(11,40)&(14,19)\\
(15,19)          &(15,22)\ensuremath{\ldots}(15,40) &(12,16)&(14,21)\ensuremath{\ldots}(14,26)\\
(15,21)\ensuremath{\ldots}(15,23)&(16,19)           &(12,18)&(14,28)\ensuremath{\ldots}(14,40)\\
(15,25)\ensuremath{\ldots}(15,40)&(16,20)           &(12,19)&(15,24)\\
(16,18)          &(16,22)           &(12,21)&(15,26)\ensuremath{\ldots}(15,31)\\
(16,20)          &(16,23)           &(12,23)&(15,33)\ensuremath{\ldots}(15,40)\\
(16,22)          &(16,26)\ensuremath{\ldots}(16,40) &(12,25)\ensuremath{\ldots}(12,27)&(16,17)\\
(16,24)\ensuremath{\ldots}(16,40)&(17,21)           &(12,29)\ensuremath{\ldots}(12,40)&(16,19)\\
(17,20)          &(17,24)\ensuremath{\ldots}(17,40) &(13,17)\ensuremath{\ldots}(13,19)&(16,22)\\
(17,22)\ensuremath{\ldots}(17,24)&(18,23)           &(13,23)&(16,26)\\
(17,26)\ensuremath{\ldots}(17,40)&(18,26)\ensuremath{\ldots}(18,40) &(13,25)&(16,28)\\
(18,21)          &(19,24)           &(13,27)&(16,30)\\
(18,23)\ensuremath{\ldots}(18,40)&(19,25)           &(13,28)&(16,31)\\
(19,24)\ensuremath{\ldots}(19,26)&(19,27)           &(13,30)\ensuremath{\ldots}(13,40)&(16,33)\\
(19,29)\ensuremath{\ldots}(19,40)&(19,29)\ensuremath{\ldots}(19,40) &(14,17)&(16,35)\ensuremath{\ldots}(16,40)\\
(20,23)          &(20,24)\ensuremath{\ldots}(20,26) &(14,20)\ensuremath{\ldots}(14,40)&(17,21)\\
(20,25)\ensuremath{\ldots}(20,40)&(20,28)\ensuremath{\ldots}(20,40) &(15,18)&(17,23)\ensuremath{\ldots}(17,25)\\
(21,24)          &(21,23)           &(15,19)&(17,27)\ensuremath{\ldots}(17,40)\\
(21,26)\ensuremath{\ldots}(21,40)&(21,25)           &(15,23)&(18,22)\\
(22,25)          &(21,27)\ensuremath{\ldots}(21,40) &(15,25)&(18,23)\\
(22,29)\ensuremath{\ldots}(22,40)&(22,27)\ensuremath{\ldots}(22,40) &(15,27)&(18,25)\ensuremath{\ldots}(18,40)\\
(23,26)          &(23,30)\ensuremath{\ldots}(23,40) &(15,30)&(19,24)\\
(23,28)          &(24,27)           &(15,34)&(19,25)\\
(23,31)\ensuremath{\ldots}(23,40)&(24,31)\ensuremath{\ldots}(24,40) &(15,36)\ensuremath{\ldots}(15,40)&(19,27)\ensuremath{\ldots}(19,29)\\
(24,27)\ensuremath{\ldots}(24,30)&(25,29)           &(16,20)&(19,31)\ensuremath{\ldots}(19,40)\\
(24,32)\ensuremath{\ldots}(24,40)&(25,32)\ensuremath{\ldots}(25,40) &(16,22)\ensuremath{\ldots}(16,25)&(20,26)\\
(25,28)          &(26,30)\ensuremath{\ldots}(26,40) &(16,27)&(20,28)\\
(25,30)          &(27,33)\ensuremath{\ldots}(27,40) &(16,28)&(20,30)\ensuremath{\ldots}(20,40)\\
(25,31)          &(28,34)\ensuremath{\ldots}(28,40) &(16,30)\ensuremath{\ldots}(16,40)&(21,26)\ensuremath{\ldots}(21,40)\\
(25,33)\ensuremath{\ldots}(25,40)&(29,35)\ensuremath{\ldots}(29,40) &(17,21)\ensuremath{\ldots}(17,40)&(22,27)\\
(26,29)          &(30,35)\ensuremath{\ldots}(30,40) &(18,20)&(22,29)\\
(26,30)          &(31,35)           &(18,22)&(22,31)\ensuremath{\ldots}(22,40)\\
(26,32)          &(31,37)\ensuremath{\ldots}(31,40) &(18,24)&(23,26)\\
(26,34)\ensuremath{\ldots}(26,40)&(32,34)           &(18,28)&(23,27)\\
(27,31)\ensuremath{\ldots}(27,40)&(32,36)\ensuremath{\ldots}(32,40) &(18,31)&(23,29)\ensuremath{\ldots}(23,31)\\
(28,32)          &(33,39)           &(18,33)\ensuremath{\ldots}(18,40)&(23,33)\ensuremath{\ldots}(23,40)\\
(28,34)\ensuremath{\ldots}(28,40)&(33,40)           &(19,22)&(24,33)\ensuremath{\ldots}(24,40)\\
(29,31)          &(34,38)\ensuremath{\ldots}(34,40) &(19,24)\ensuremath{\ldots}(19,40)&(25,30)\\
(29,33)\ensuremath{\ldots}(29,40)&(35,39)           &(20,23)\ensuremath{\ldots}(20,40)&(25,31)\\
(30,35)          &(35,40)           &(21,27)\ensuremath{\ldots}(21,40)&(25,34)\ensuremath{\ldots}(25,40)\\
(30,36)          &(36,40)           &(22,25)\ensuremath{\ldots}(22,40)&(26,31)\ensuremath{\ldots}(26,40)\\
(30,38)\ensuremath{\ldots}(30,40)&                  &(23,26)&(27,31)\ensuremath{\ldots}(27,40)\\
(31,34)          &                  &(23,28)\ensuremath{\ldots}(23,40)&(28,33)\\
(31,36)\ensuremath{\ldots}(31,40)&                  &(24,27)&(28,34)\\
(32,35)          &                  &(24,29)&(28,37)\ensuremath{\ldots}(28,40)\\
(32,36)          &                  &(24,30)&(29,35)\ensuremath{\ldots}(29,40)\\
(32,38)\ensuremath{\ldots}(32,40)&                  &(24,32)\ensuremath{\ldots}(24,40)&(30,33)\ensuremath{\ldots}(30,35)\\
(33,36)\ensuremath{\ldots}(33,40)&                  &(25,28)\ensuremath{\ldots}(25,40)&(30,37)\ensuremath{\ldots}(30,40)\\
(34,40)          &                  &(26,31)\ensuremath{\ldots}(26,40)&(31,37)\ensuremath{\ldots}(31,40)\\
(35,37)\ensuremath{\ldots}(35,40)&                  &(27,31)\ensuremath{\ldots}(27,40)&(32,37)\\
(37,39)          &                  &(28,30)&(32,39)\\
(37,40)          &                  &(28,32)\ensuremath{\ldots}(28,40)&(32,40)\\
&&                                   (29,34)\ensuremath{\ldots}(29,40)&(33,39)\\
&&                                   (30,33)&(33,40)\\
&&                                   (30,35)\ensuremath{\ldots}(30,40)&(34,39)\\
&&                                   (31,34)\ensuremath{\ldots}(31,40)&(34,40)\\
&&                                   (32,37)\ensuremath{\ldots}(32,40)&(36,40)\\
&&                                   (33,36)&\\
&&                                   (33,38)\ensuremath{\ldots}(33,40)&\\ &&                                   (34,39)&\\
&&                                   (34,40)&\\
&&                                   (35,40)&\\
&&                                   (36,40)&\\
&&                                   (37,40)&\\
\hline
\end{longtable}

\newpage
\begin{longtable}{|l|l|l|l|}
\hline
\multicolumn{4}{|c|}{GA-BISON}\\
\multicolumn{2}{|c|}{No Obstacles}&\multicolumn{2}{|c|}{10 Scatterers}\\
\multicolumn{1}{|c}{No Noise}&\multicolumn{1}{|c}{Noisy}&\multicolumn{1}{|c}{No Noise}&\multicolumn{1}{|c|}{Noisy}\\
\hline
(1,5)&(1,9)&(1,5)&(1,8)\\
(1,6)&(1,14)&(1,7)&(1,10)\\
(1,9)&(1,15)&(1,9)\ensuremath{\ldots}(1,12)&(1,11)\\
(1,10)&(1,17)&(1,14)&(1,18)\\
(1,12)\ensuremath{\ldots}(1,14)&(1,21)&(1,15)&(1,20)\\
(1,17)\ensuremath{\ldots}(1,20)&(1,25)\ensuremath{\ldots}(1,40)&(1,19)\ensuremath{\ldots}(1,21)&(1,21)\\
(1,22)\ensuremath{\ldots}(1,40)&(2,5)&(1,23)\ensuremath{\ldots}(1,26)&(1,23)\ensuremath{\ldots}(1,25)\\
(2,5)&(2,9)\ensuremath{\ldots}(2,11)&(1,28)\ensuremath{\ldots}(1,40)&(1,27)\ensuremath{\ldots}(1,40)\\
(2,6)&(2,15)&(2,6)&(2,8)\\
(2,9)&(2,17)&(2,8)&(2,10)\ensuremath{\ldots}(2,18)\\
(2,10)&(2,19)&(2,11)&(2,20)\\
(2,12)\ensuremath{\ldots}(2,15)&(2,27)\ensuremath{\ldots}(2,29)&(2,13)&(2,21)\\
(2,17)\ensuremath{\ldots}(2,19)&(2,31)\ensuremath{\ldots}(2,40)&(2,16)\ensuremath{\ldots}(2,22)&(2,23)\ensuremath{\ldots}(2,29)\\
(2,22)\ensuremath{\ldots}(2,25)&(3,9)&(2,24)\ensuremath{\ldots}(2,40)&(2,31)\ensuremath{\ldots}(2,40)\\
(2,27)\ensuremath{\ldots}(2,29)&(3,15)&(3,6)&(3,4)\\
(2,31)\ensuremath{\ldots}(2,40)&(3,21)&(3,11)&(3,9)\\
(3,9)\ensuremath{\ldots}(3,15)&(3,24)\ensuremath{\ldots}(3,40)&(3,13)&(3,14)\\
(3,17)&(4,7)&(3,15)&(3,16)\ensuremath{\ldots}(3,40)\\
(3,19)&(4,9)&(3,16)&(4,8)\\
(3,22)\ensuremath{\ldots}(3,24)&(4,15)\ensuremath{\ldots}(4,17)&(3,18)\ensuremath{\ldots}(3,40)&(4,10)\\
(3,27)\ensuremath{\ldots}(3,29)&(4,24)&(4,6)&(4,11)\\
(3,31)\ensuremath{\ldots}(3,40)&(4,26)\ensuremath{\ldots}(4,28)&(4,8)&(4,13)\\
(4,17)&(4,30)&(4,16)&(4,14)\\
(4,22)&(4,32)\ensuremath{\ldots}(4,34)&(4,18)\ensuremath{\ldots}(4,22)&(4,17)\\
(4,24)&(4,36)\ensuremath{\ldots}(4,40)&(4,25)\ensuremath{\ldots}(4,40)&(4,18)\\
(4,26)&(5,10)&(5,11)&(4,20)\\
(4,27)&(5,12)\ensuremath{\ldots}(5,14)&(5,12)&(4,23)\ensuremath{\ldots}(4,27)\\
(4,30)&(5,16)&(5,16)\ensuremath{\ldots}(5,19)&(4,29)\ensuremath{\ldots}(4,40)\\
(4,32)\ensuremath{\ldots}(4,40)&(5,18)\ensuremath{\ldots}(5,23)&(5,22)&(5,14)\\
(5,16)&(5,25)&(5,25)\ensuremath{\ldots}(5,40)&(5,17)\\
(5,18)\ensuremath{\ldots}(5,21)&(5,26)&(6,12)\ensuremath{\ldots}(6,15)&(5,18)\\
(5,25)&(5,28)\ensuremath{\ldots}(5,40)&(6,21)&(5,20)\ensuremath{\ldots}(5,40)\\
(5,26)&(6,9)\ensuremath{\ldots}(6,11)&(6,24)&(6,11)\\
(5,28)&(6,15)&(6,25)&(6,16)\ensuremath{\ldots}(6,40)\\
(5,30)\ensuremath{\ldots}(5,32)&(6,17)\ensuremath{\ldots}(6,22)&(6,28)&(7,14)\\
(5,34)\ensuremath{\ldots}(5,40)&(6,25)\ensuremath{\ldots}(6,40)&(6,30)\ensuremath{\ldots}(6,40)&(7,17)\\
(6,11)&(7,10)&(7,8)&(7,18)\\
(6,14)&(7,14)&(7,11)&(7,20)\\
(6,15)&(7,17)\ensuremath{\ldots}(7,21)&(7,13)&(7,21)\\
(6,17)&(7,23)&(7,16)&(7,23)\ensuremath{\ldots}(7,40)\\
(6,18)&(7,25)\ensuremath{\ldots}(7,40)&(7,18)&(8,12)\\
(6,20)\ensuremath{\ldots}(6,40)&(8,13)&(7,19)&(8,14)\\
(7,11)&(8,14)&(7,22)&(8,16)\\
(7,14)&(8,18)\ensuremath{\ldots}(8,20)&(7,25)\ensuremath{\ldots}(7,40)&(8,19)\ensuremath{\ldots}(8,23)\\
(7,22)&(8,22)&(8,10)&(8,25)\ensuremath{\ldots}(8,40)\\
(7,24)&(8,23)&(8,11)&(9,18)\\
(7,26)\ensuremath{\ldots}(7,40)&(8,25)&(8,14)&(9,20)\\
(8,11)&(8,26)&(8,15)&(9,21)\\
(8,20)\ensuremath{\ldots}(8,22)&(8,28)\ensuremath{\ldots}(8,40)&(8,17)\ensuremath{\ldots}(8,21)&(9,25)\\
(8,24)\ensuremath{\ldots}(8,40)&(9,13)&(8,23)\ensuremath{\ldots}(8,26)&(9,27)\\
(9,16)&(9,14)&(8,29)\ensuremath{\ldots}(8,40)&(9,29)\ensuremath{\ldots}(9,40)\\
(9,20)&(9,18)\ensuremath{\ldots}(9,26)&(9,15)&(10,19)\ensuremath{\ldots}(10,21)\\
(9,21)&(9,29)\ensuremath{\ldots}(9,33)&(9,21)&(10,23)\\
(9,26)&(9,35)\ensuremath{\ldots}(9,40)&(9,22)&(10,25)\ensuremath{\ldots}(10,33)\\
(9,29)\ensuremath{\ldots}(9,40)&(10,13)&(9,24)\ensuremath{\ldots}(9,26)&(10,35)\ensuremath{\ldots}(10,40)\\
(10,18)\ensuremath{\ldots}(10,21)&(10,15)\ensuremath{\ldots}(10,17)&(9,28)\ensuremath{\ldots}(9,40)&(11,19)\\
(10,23)\ensuremath{\ldots}(10,26)&(10,20)&(10,16)&(11,21)\ensuremath{\ldots}(11,23)\\
(10,28)\ensuremath{\ldots}(10,40)&(10,24)&(10,18)&(11,25)\ensuremath{\ldots}(11,33)\\
(11,18)&(10,25)&(10,19)&(11,35)\ensuremath{\ldots}(11,40)\\
(11,22)\ensuremath{\ldots}(11,24)&(10,27)&(10,22)&(12,18)\\
(11,27)\ensuremath{\ldots}(11,29)&(10,28)&(10,25)\ensuremath{\ldots}(10,40)&(12,20)\\
(11,33)\ensuremath{\ldots}(11,40)&(10,30)\ensuremath{\ldots}(10,40)&(11,29)&(12,21)\\
(12,21)&(11,13)&(11,31)\ensuremath{\ldots}(11,40)&(12,25)\ensuremath{\ldots}(12,40)\\
(12,25)&(11,28)&(12,15)&(13,19)\ensuremath{\ldots}(13,23)\\
(12,27)&(11,30)\ensuremath{\ldots}(11,33)&(12,18)&(13,25)\ensuremath{\ldots}(13,27)\\
(12,38)\ensuremath{\ldots}(12,40)&(11,35)\ensuremath{\ldots}(11,40)&(12,19)&(13,29)\ensuremath{\ldots}(13,40)\\
(13,20)\ensuremath{\ldots}(13,40)&(12,17)&(12,21)\ensuremath{\ldots}(12,23)&(14,29)\\
(14,20)&(12,20)&(12,25)\ensuremath{\ldots}(12,40)&(14,32)\\
(14,21)&(12,21)&(13,20)\ensuremath{\ldots}(13,26)&(14,34)\ensuremath{\ldots}(14,40)\\
(14,30)\ensuremath{\ldots}(14,32)&(12,24)\ensuremath{\ldots}(12,28)&(13,28)\ensuremath{\ldots}(13,40)&(15,19)\\
(14,34)\ensuremath{\ldots}(14,40)&(12,30)\ensuremath{\ldots}(12,40)&(14,16)&(15,21)\\
(15,18)&(13,17)\ensuremath{\ldots}(13,21)&(14,19)\ensuremath{\ldots}(14,40)&(15,27)\\
(15,20)&(13,25)\ensuremath{\ldots}(13,40)&(15,22)&(15,29)\ensuremath{\ldots}(15,40)\\
(15,21)&(14,21)&(15,25)\ensuremath{\ldots}(15,28)&(16,23)\\
(15,23)&(14,24)&(15,30)&(16,25)\\
(15,25)&(14,25)&(15,34)\ensuremath{\ldots}(15,40)&(16,27)\\
(15,26)&(14,27)&(16,21)\ensuremath{\ldots}(16,26)&(16,29)\ensuremath{\ldots}(16,40)\\
(15,28)\ensuremath{\ldots}(15,40)&(14,28)&(16,28)\ensuremath{\ldots}(16,40)&(17,27)\\
(16,22)&(14,30)\ensuremath{\ldots}(14,40)&(17,22)\ensuremath{\ldots}(17,40)&(17,30)\\
(16,24)\ensuremath{\ldots}(16,40)&(15,29)\ensuremath{\ldots}(15,40)&(18,23)\ensuremath{\ldots}(18,26)&(17,31)\\
(17,21)&(16,25)&(18,28)\ensuremath{\ldots}(18,40)&(17,33)\\
(17,22)&(16,27)&(19,28)&(17,35)\ensuremath{\ldots}(17,40)\\
(17,24)&(16,29)\ensuremath{\ldots}(16,32)&(19,30)&(18,21)\\
(17,26)&(16,34)\ensuremath{\ldots}(16,40)&(19,31)&(18,26)\ensuremath{\ldots}(18,28)\\
(17,27)&(17,29)\ensuremath{\ldots}(17,40)&(19,33)\ensuremath{\ldots}(19,40)&(18,30)\\
(17,29)\ensuremath{\ldots}(17,40)&(18,21)&(20,25)&(18,31)\\
(18,27)\ensuremath{\ldots}(18,40)&(18,24)&(20,26)&(18,33)\\
(19,22)&(18,28)&(20,28)&(18,35)\ensuremath{\ldots}(18,40)\\
(19,24)\ensuremath{\ldots}(19,40)&(18,30)\ensuremath{\ldots}(18,34)&(20,30)&(19,21)\\
(20,27)&(18,36)\ensuremath{\ldots}(18,40)&(20,31)&(19,23)\ensuremath{\ldots}(19,27)\\
(20,29)&(19,27)\ensuremath{\ldots}(19,40)&(20,33)\ensuremath{\ldots}(20,40)&(19,29)\ensuremath{\ldots}(19,40)\\
(20,30)&(20,27)&(21,30)&(20,26)\\
(20,32)\ensuremath{\ldots}(20,40)&(20,28)&(21,34)\ensuremath{\ldots}(21,40)&(20,27)\\
(21,24)&(20,35)&(22,29)\ensuremath{\ldots}(22,40)&(20,30)\\
(21,27)\ensuremath{\ldots}(21,29)&(20,37)&(23,27)&(20,37)\ensuremath{\ldots}(20,40)\\
(21,32)\ensuremath{\ldots}(21,40)&(20,39)&(23,28)&(21,29)\\
(22,25)&(20,40)&(23,30)&(21,32)\\
(22,26)&(21,35)\ensuremath{\ldots}(21,40)&(23,31)&(21,34)\ensuremath{\ldots}(21,40)\\
(22,28)&(22,25)\ensuremath{\ldots}(22,40)&(23,33)\ensuremath{\ldots}(23,40)&(22,27)\ensuremath{\ldots}(22,29)\\
(22,30)\ensuremath{\ldots}(22,32)&(23,28)\ensuremath{\ldots}(23,40)&(24,30)&(22,31)\ensuremath{\ldots}(22,40)\\
(22,34)&(24,29)&(24,33)\ensuremath{\ldots}(24,40)&(23,31)\ensuremath{\ldots}(23,40)\\
(22,36)\ensuremath{\ldots}(22,40)&(24,31)\ensuremath{\ldots}(24,40)&(25,40)&(24,30)\ensuremath{\ldots}(24,40)\\
(23,29)\ensuremath{\ldots}(23,40)&(25,32)&(26,33)&(25,31)\\
(24,30)&(25,36)\ensuremath{\ldots}(25,40)&(26,34)&(25,33)\ensuremath{\ldots}(25,40)\\
(24,32)&(26,31)&(26,36)\ensuremath{\ldots}(26,40)&(26,32)\\
(24,34)&(26,35)\ensuremath{\ldots}(26,40)&(27,30)\ensuremath{\ldots}(27,40)&(26,34)\ensuremath{\ldots}(26,36)\\
(24,36)\ensuremath{\ldots}(24,40)&(27,32)&(28,33)\ensuremath{\ldots}(28,40)&(26,38)\ensuremath{\ldots}(26,40)\\
(25,30)&(27,33)&(29,33)\ensuremath{\ldots}(29,40)&(27,36)\\
(25,32)\ensuremath{\ldots}(25,40)&(27,35)\ensuremath{\ldots}(27,40)&(30,36)\ensuremath{\ldots}(30,40)&(27,38)\ensuremath{\ldots}(27,40)\\
(26,29)&(28,35)\ensuremath{\ldots}(28,40)&(31,34)&(28,33)\ensuremath{\ldots}(28,40)\\
(26,33)\ensuremath{\ldots}(26,40)&(29,35)\ensuremath{\ldots}(29,40)&(31,35)&(29,35)\\
(27,31)&(30,40)&(31,37)\ensuremath{\ldots}(31,40)&(29,37)\\
(27,34)\ensuremath{\ldots}(27,40)&(31,37)\ensuremath{\ldots}(31,40)&(32,35)\ensuremath{\ldots}(32,40)&(29,39)\\
(28,32)\ensuremath{\ldots}(28,40)&(32,39)&(33,39)&(29,40)\\
(29,36)\ensuremath{\ldots}(29,40)&(32,40)&(33,40)&(30,34)\ensuremath{\ldots}(30,40)\\
(30,33)\ensuremath{\ldots}(30,40)&(33,39)&(35,40)&(31,40)\\
(31,33)&(33,40)&&(32,36)\ensuremath{\ldots}(32,40)\\
(31,35)\ensuremath{\ldots}(31,40)&(34,37)\ensuremath{\ldots}(34,40)&&(33,40)\\
(32,35)&(35,39)&&(34,39)\\
(32,37)\ensuremath{\ldots}(32,40)&(35,40)&&(34,40)\\
(33,36)&(36,40)&&(37,40)\\
(33,38)\ensuremath{\ldots}(33,40)&&&\\
(35,38)\ensuremath{\ldots}(35,40)&&&\\
(36,39)&&&\\
(36,40)&&&\\
(37,40)&&&\\
\hline
\end{longtable}

Significant gaps are present in the edge-centric representation of our temporal network graphs. This \textit{``banding''} is inherently due to the associative nature of the two algorithms; in both cases, nodes are far more likely to connect with nodes sharing similar deployment times, due to the spatial connectivity limitations. In other words, nodes stay close to their initial neighbors, and rarely do nodes \textit{``sneak''} into other parts of the network. Investigating the gaps also highlights an issue with this representation: since edges are not double counted ((1,40) can be found on the y axis, while (40,1) is omitted), the gaps are biased towards being narrower towards the top of the figure.

\newpage
\newpage
\section*{SOM D}
\subsection*{BISON-Voronoi, No Obstacles, No Noise}
\begin{figure}[h!]
    \centering
    \includegraphics[width = 0.65\hsize]{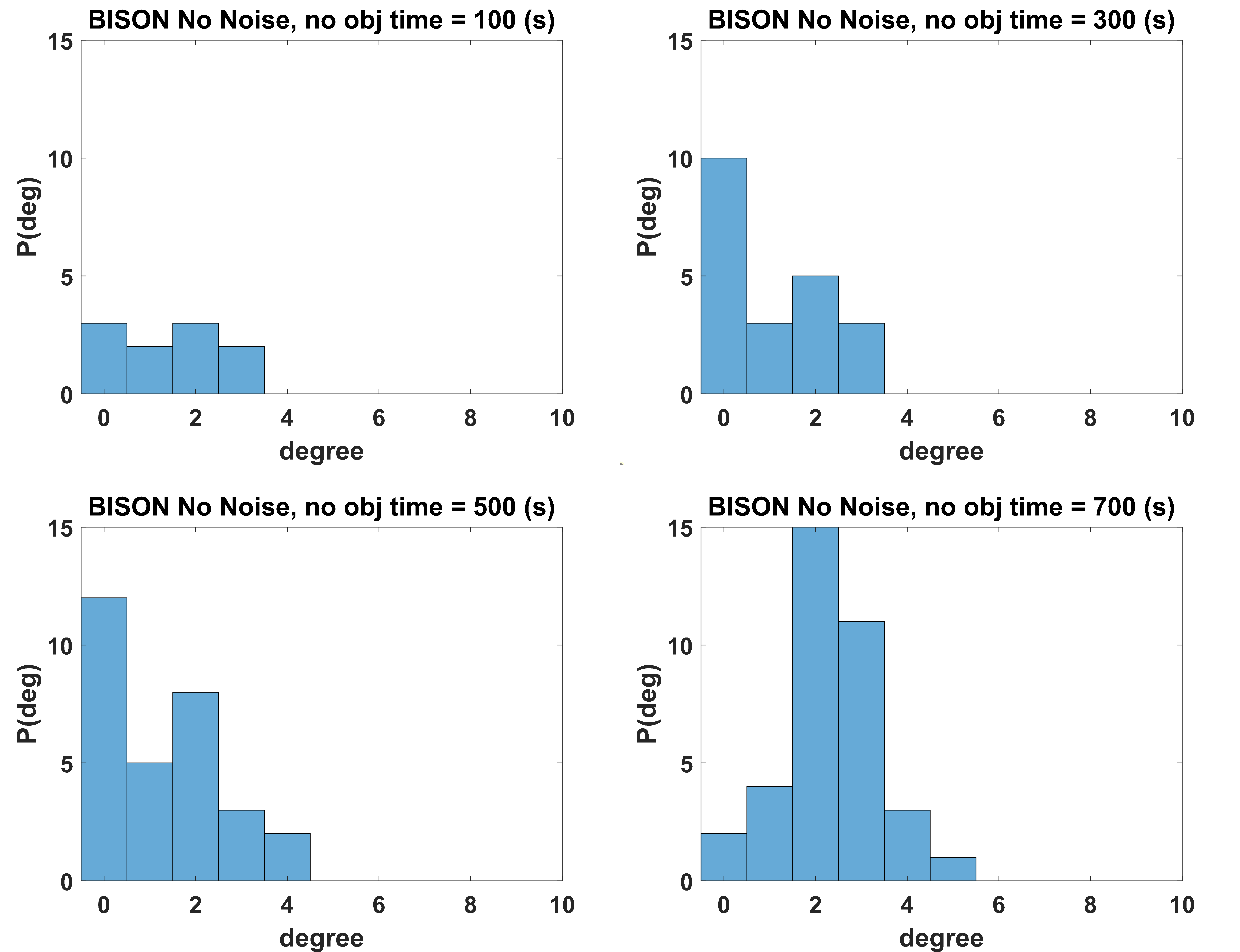}
\end{figure}

\subsection*{BISON-Voronoi, No Obstacles, Noisy}
\begin{figure}[h!]
    \centering
    \includegraphics[width = 0.65\hsize]{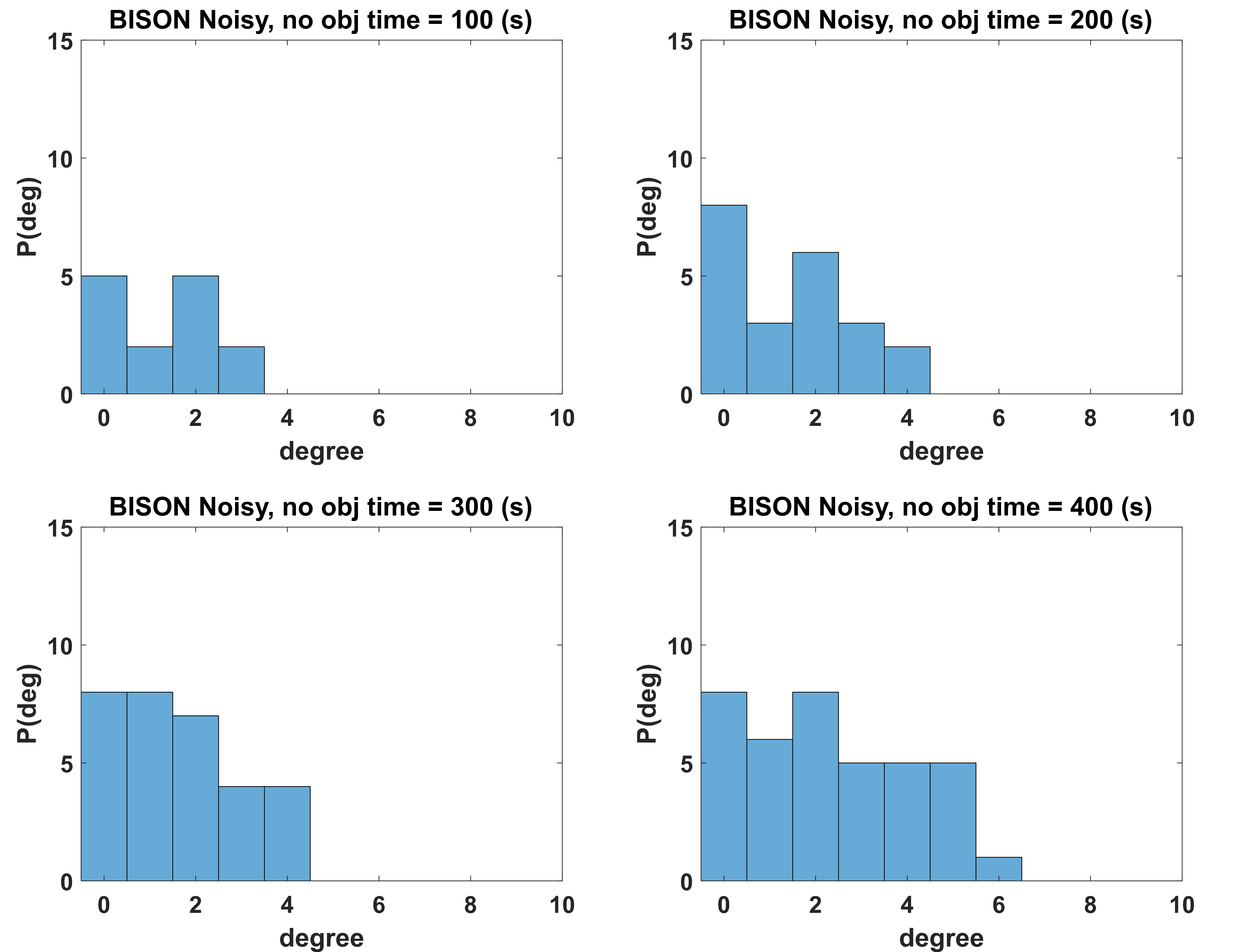}
\end{figure}
\newpage

\subsection*{BISON-Voronoi, 10 Obstacles, No Noise}
\begin{figure}[h!]
    \centering
    \includegraphics[width = 0.65\hsize]{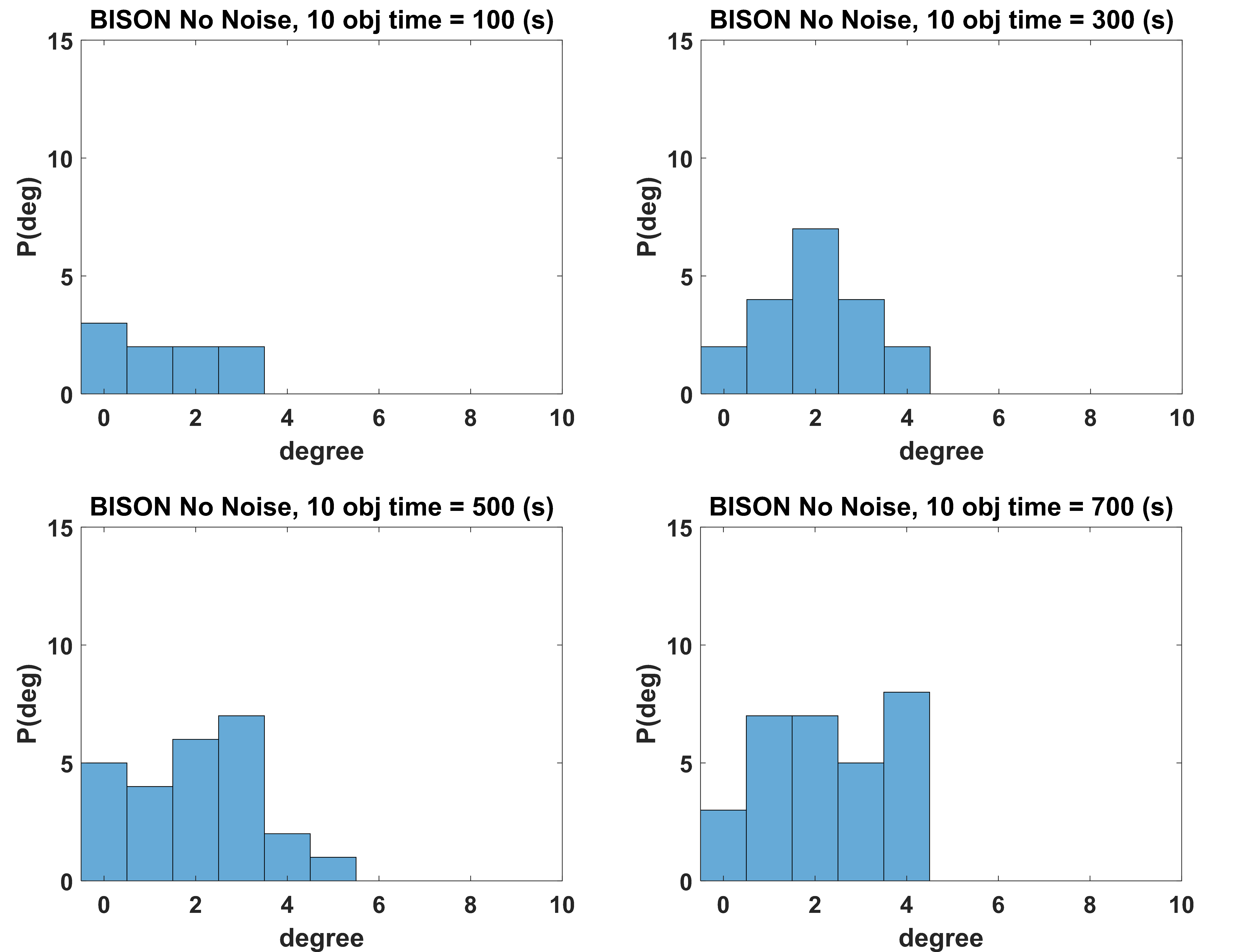}
\end{figure}

\subsection*{BISON-Voronoi, 10 Obstacles, Noisy}
\begin{figure}[h!]
    \centering
    \includegraphics[width = 0.65\hsize]{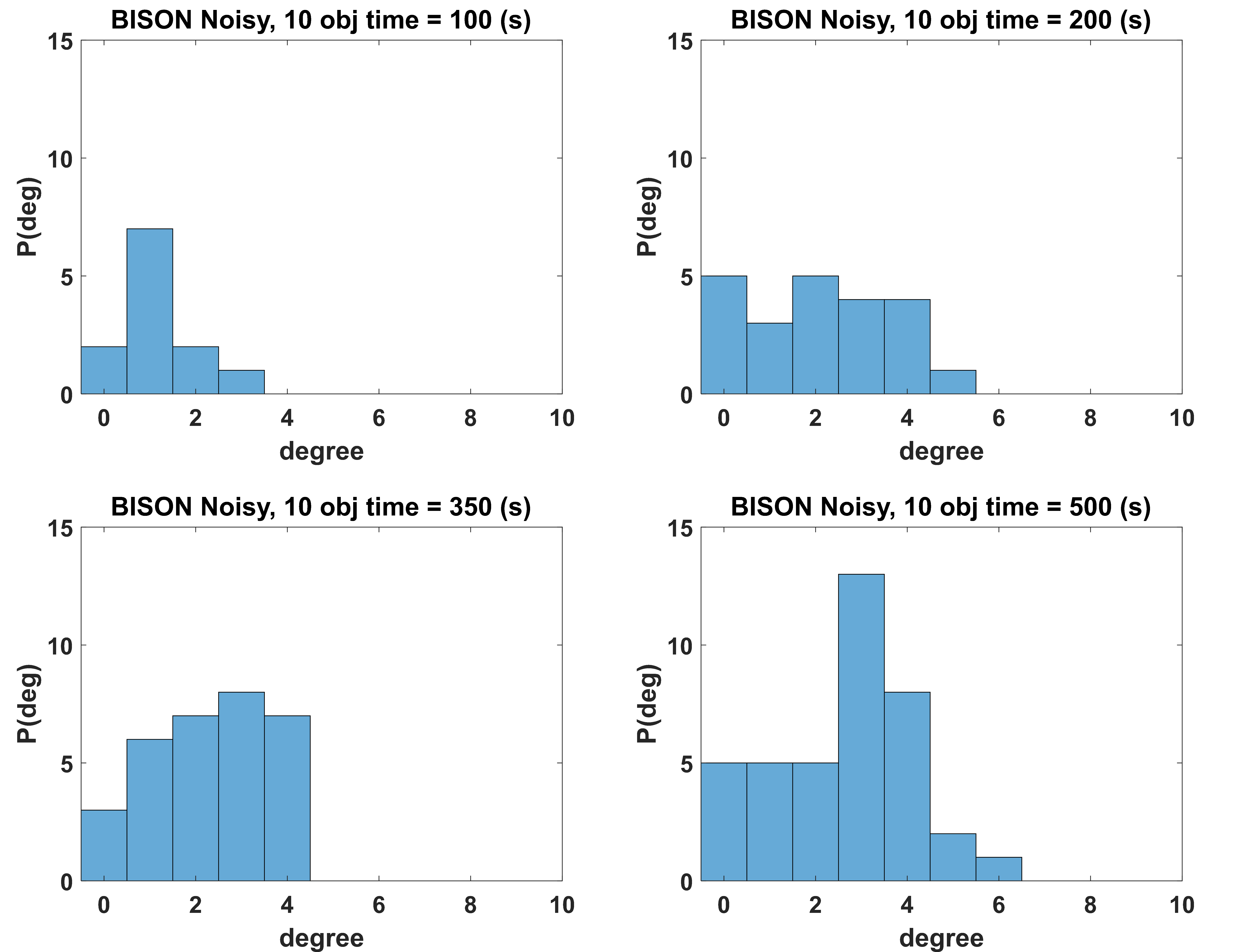}
\end{figure}
\newpage

\subsection*{GA-Voronoi, No Obstacles, No Noise}
\begin{figure}[h!]
    \centering
    \includegraphics[width = 0.65\hsize]{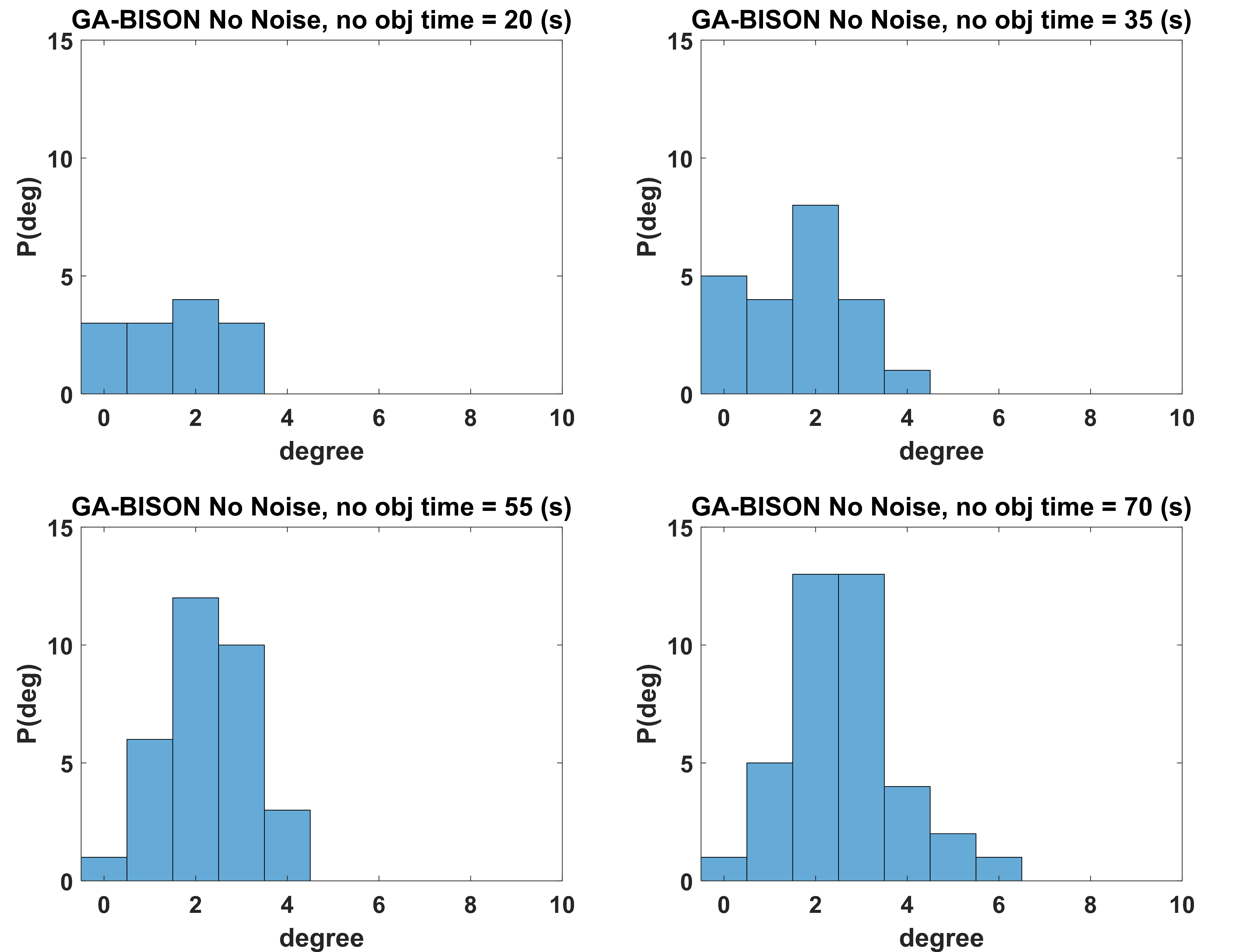}
\end{figure}

\subsection*{GA-Voronoi, No Obstacles, Noisy}
\begin{figure}[h!]
    \centering
    \includegraphics[width = 0.65\hsize]{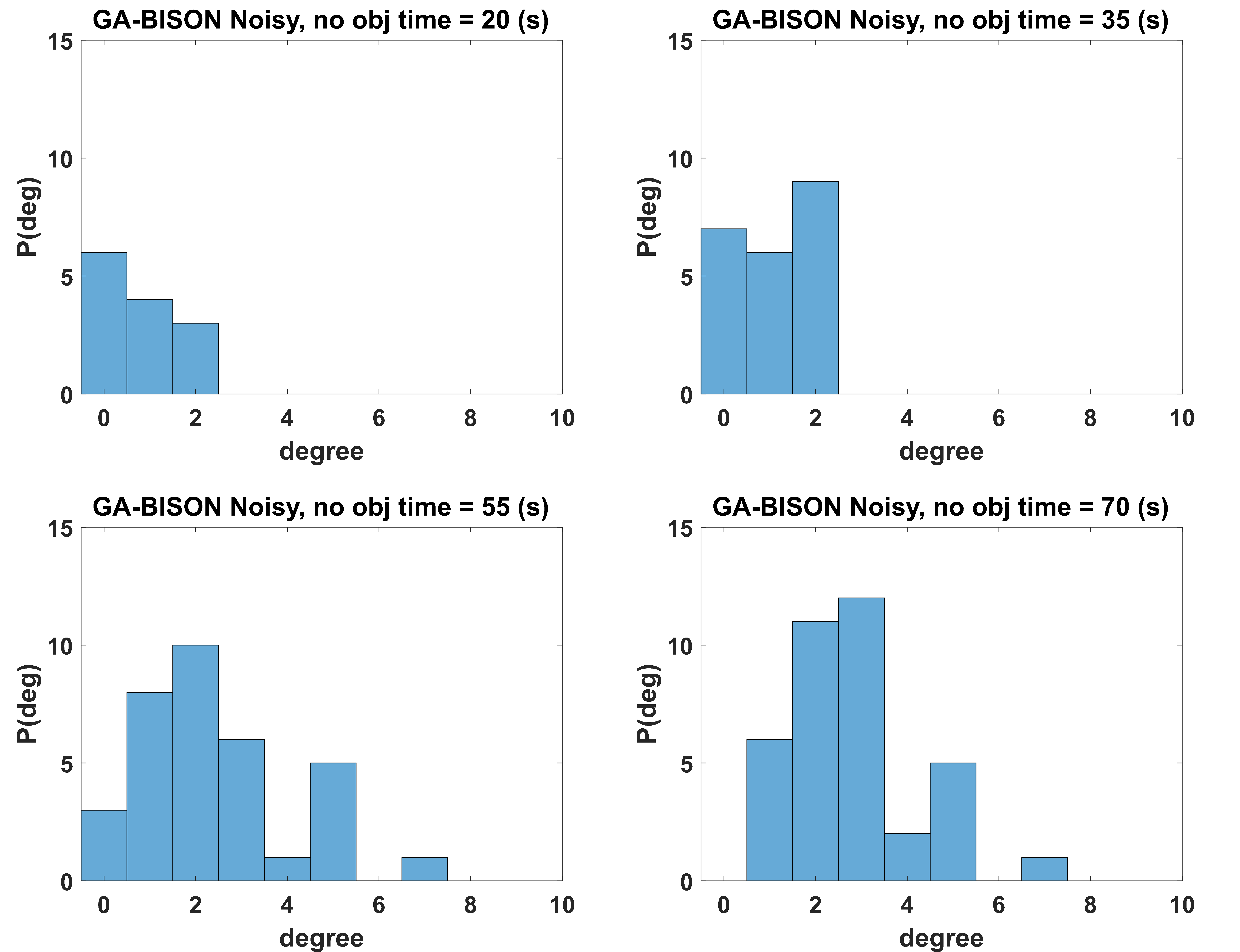}
\end{figure}
\newpage

\subsection*{GA-Voronoi, 10 Obstacles, No Noise}
\begin{figure}[h!]
    \centering
    \includegraphics[width = 0.65\hsize]{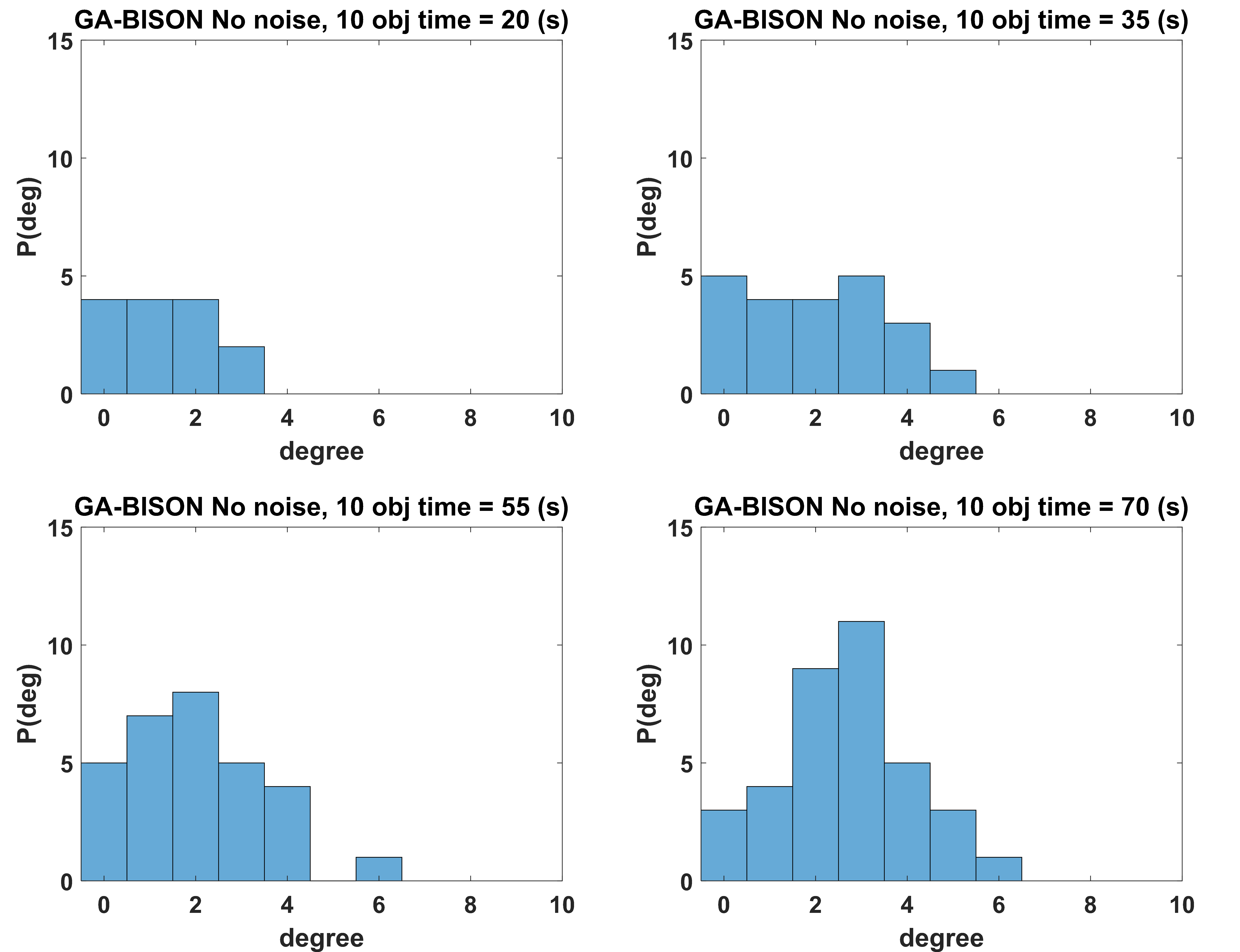}
\end{figure}

\subsection*{GA-Voronoi, 10 Obstacles, Noisy}
\begin{figure}[h!]
    \centering
    \includegraphics[width = 0.65\hsize]{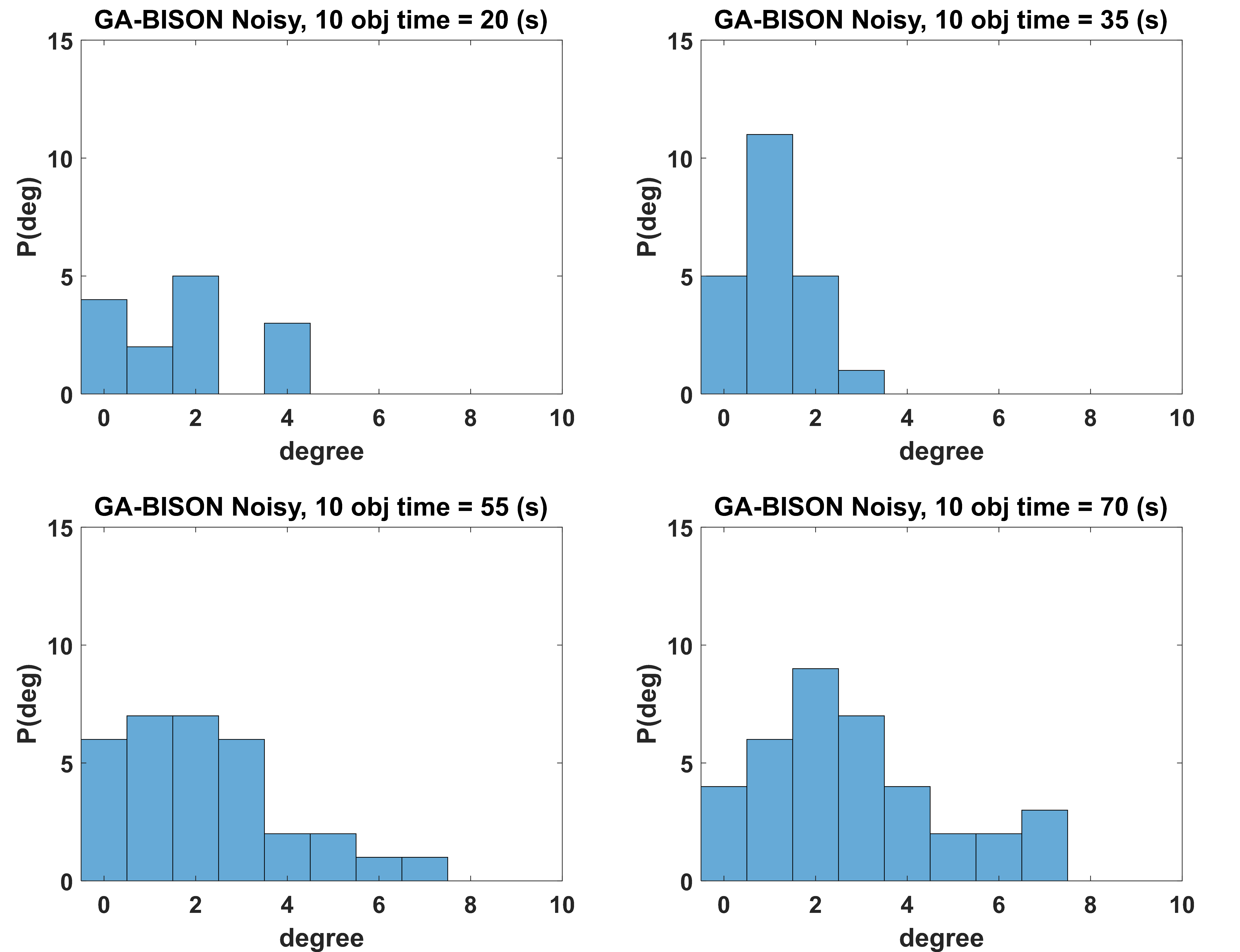}
\end{figure}

These above snapshots illustrate how the degree distributions evolve in time for the various environmental cases. We can see that the average degree typically increases with time as the network fills the space. We can also see that the distribution becomes wider over time, corresponding to the increasing regularity difference we also observed.

\newpage
\section*{SOM E}
We are displaying these here to show the variation in eigenvector centrality time traces amongst close neighbor nodes. The time trace for three nodes are plotted in each graph, and a variety of environmental conditions were used. These plots are extensions of the ones featured in Figures 6 and 7 of the main text.

\subsection*{Nodes 1, 2, and 3}
\begin{figure}[h!]
    \centering
    \includegraphics[width = \hsize]{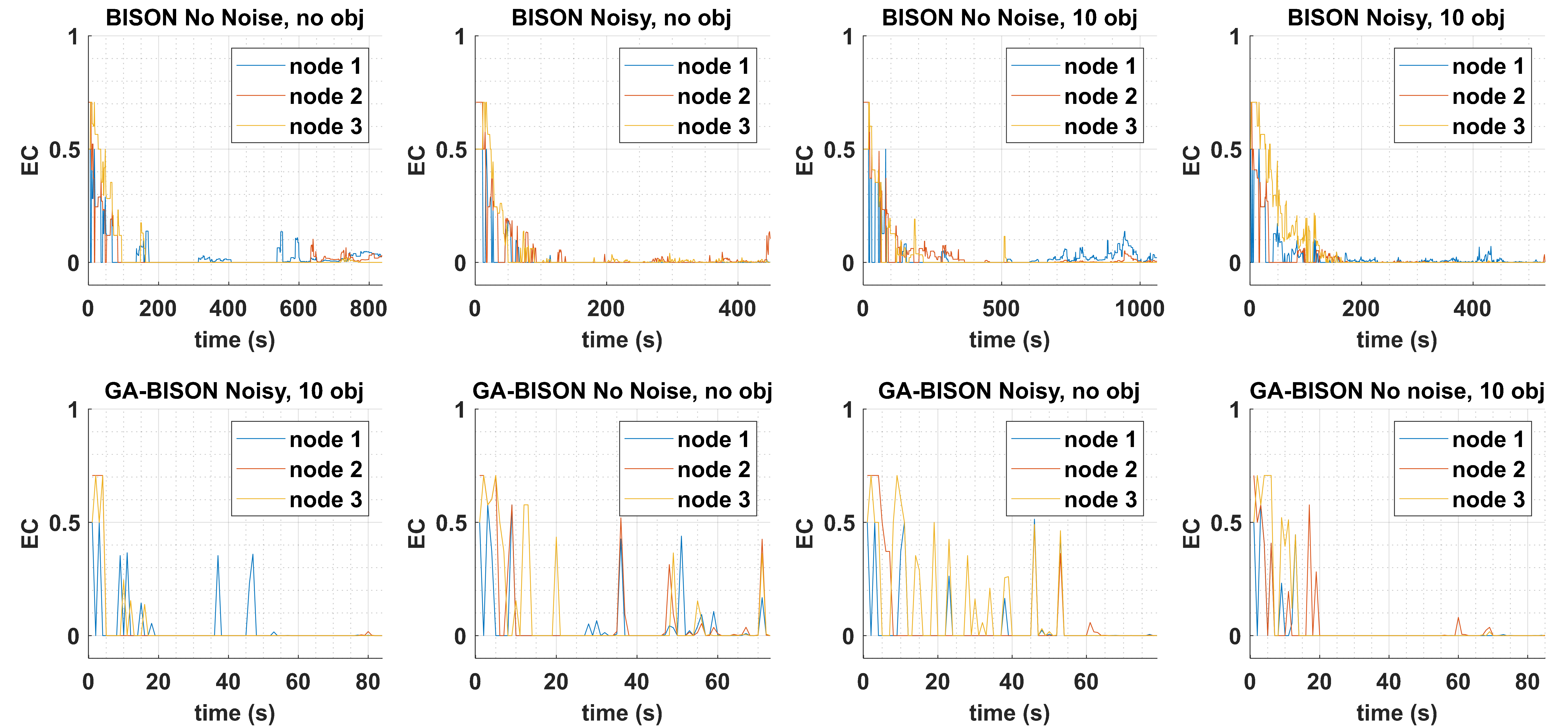}
\end{figure}

\subsection*{Nodes 4, 5, and 6}
\begin{figure}[h!]
    \centering
    \includegraphics[width = \hsize]{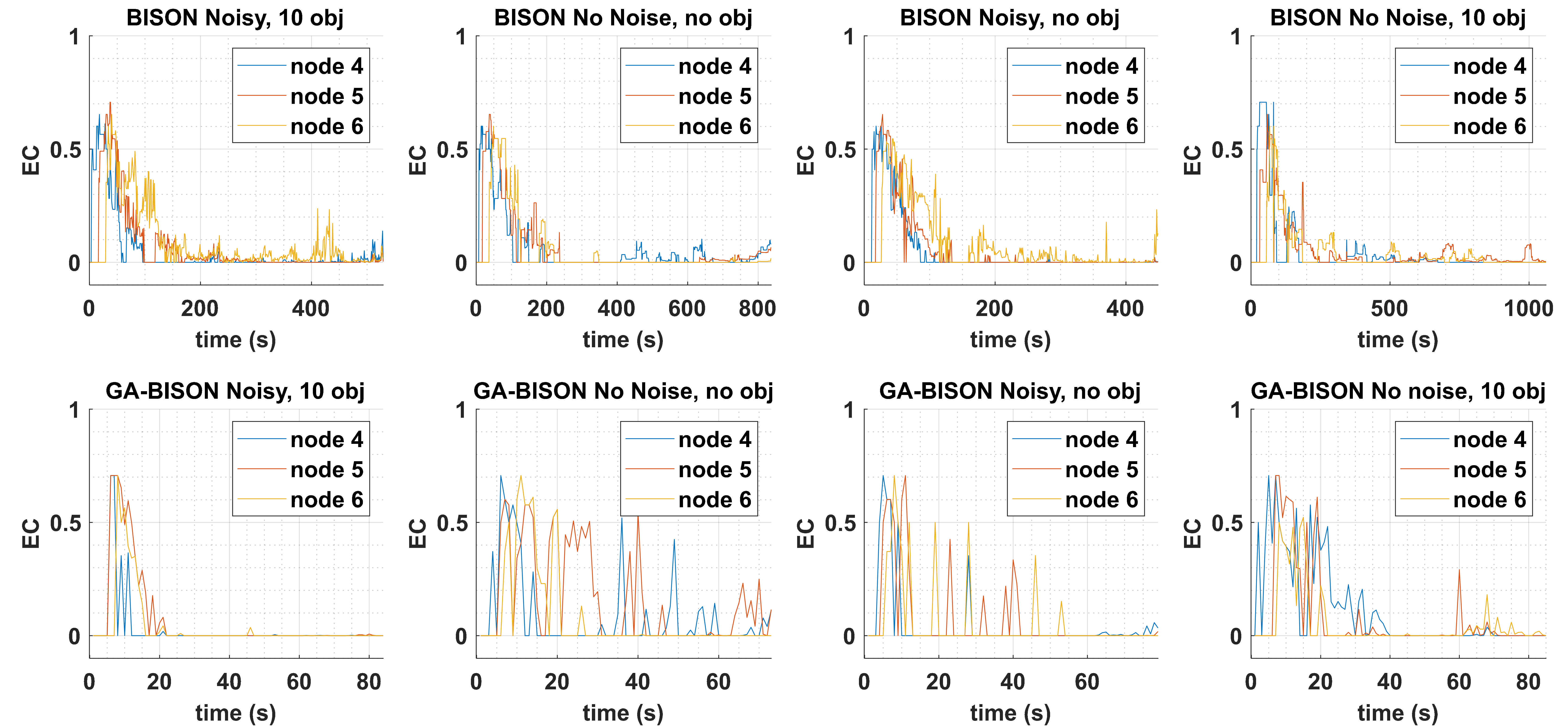}
\end{figure}
\newpage

\subsection*{Nodes 11, 12, and 13}
\begin{figure}[h!]
    \centering
    \includegraphics[width = \hsize]{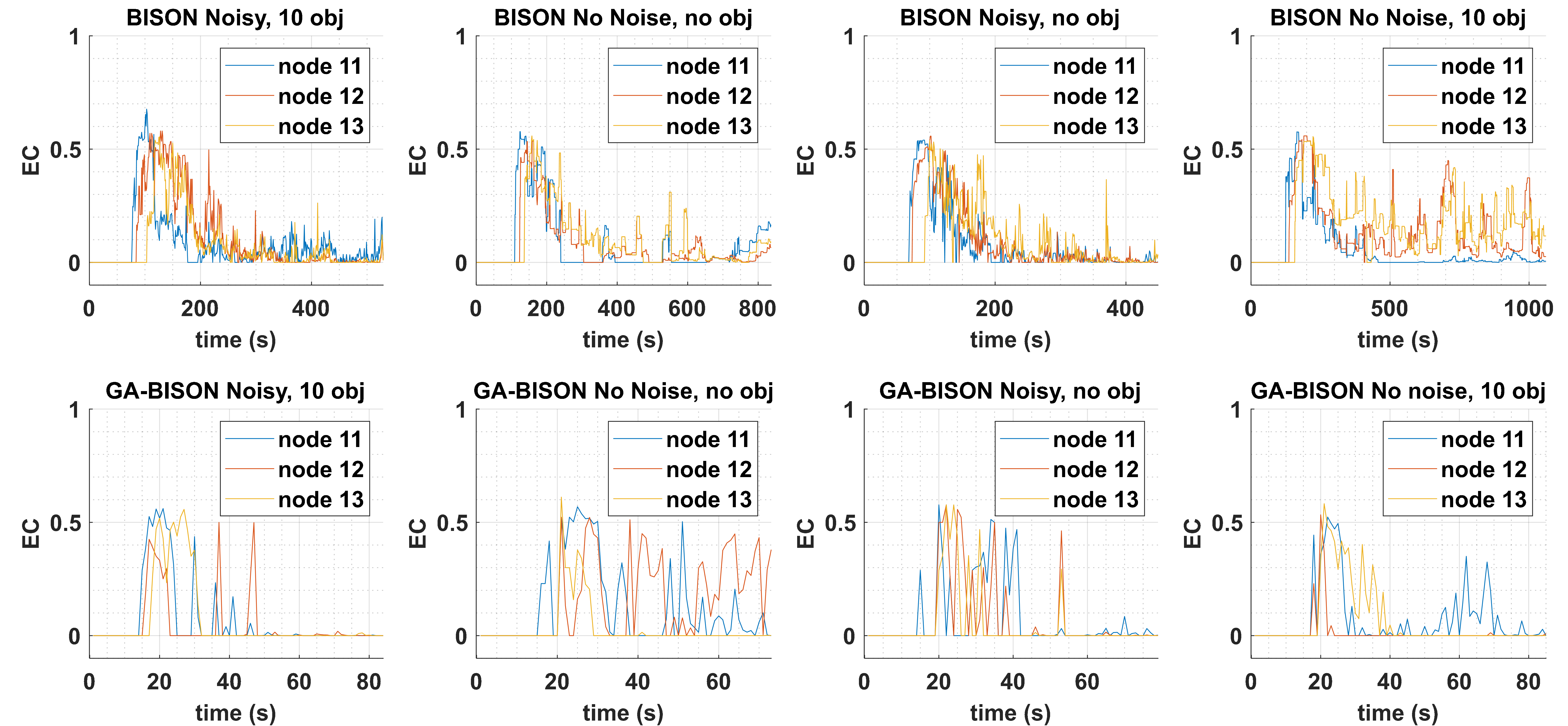}
\end{figure}

\subsection*{Nodes 19, 20, and 21}
\begin{figure}[h!]
    \centering
    \includegraphics[width = \hsize]{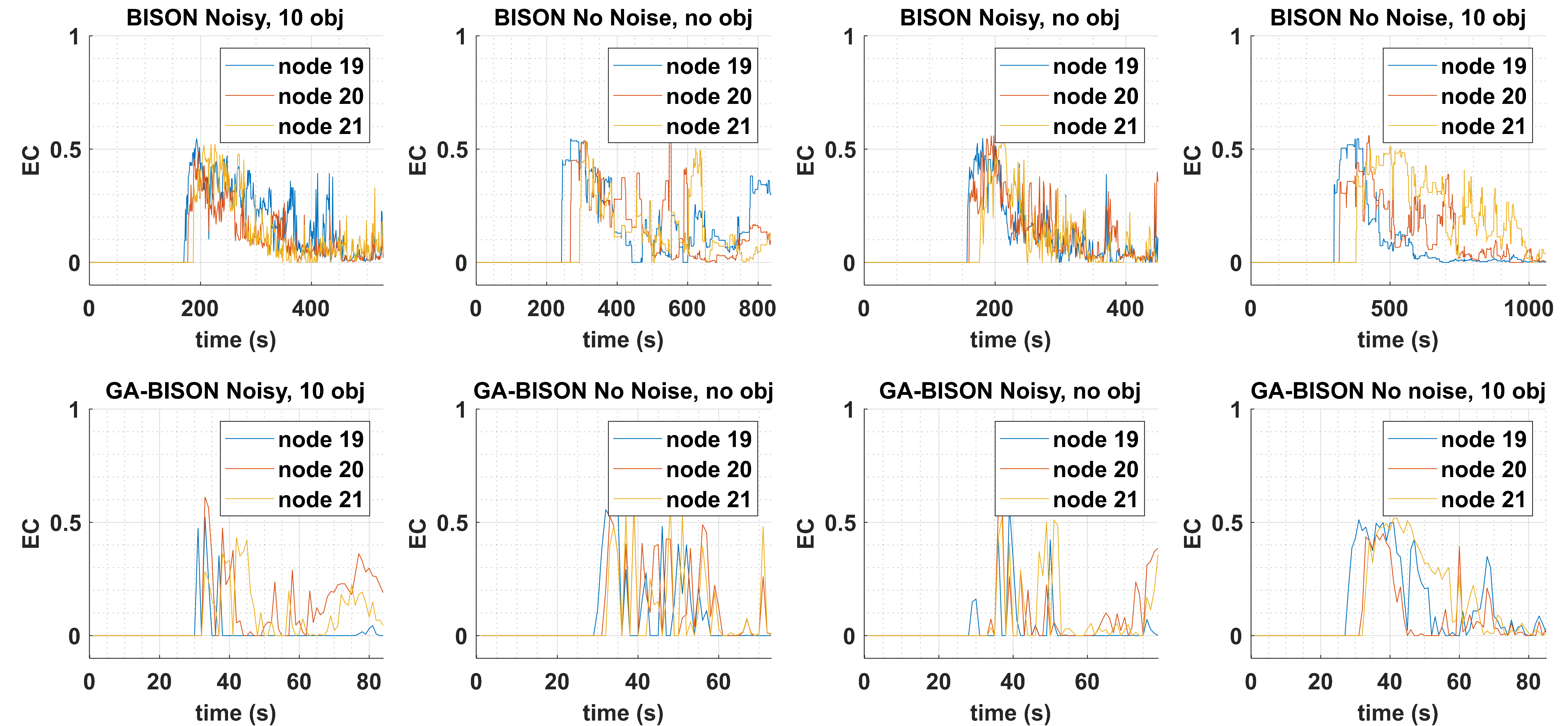}
\end{figure}

Nodes with similar deployment times exhibit qualitatively similar behavior in their eigenvector centrality scores. For BISON-Voronoi, this typically means a sharp spike when the node enters the network, followed by a fluctuating decay. These fluctuations are more pronounced in the GA-Voronoi cases, and nodes are more likely to find positions of increased importance later in their lifetimes than their BISON counterparts.

\end{document}